\documentclass[printer]{aa}

\usepackage{graphicx}
\usepackage{natbib}
\usepackage{amssymb}
\usepackage{amsmath}
\usepackage{xspace}
\usepackage[dvipsnames]{xcolor}

\usepackage[normalem]{ulem}
\usepackage[varg]{txfonts}

\newcommand{\D }{\mathrm{d}}

\newcommand{\rb}[1]{{\color{black}  #1}} 
\newcommand{\rbn}[1]{{\color{black}  #1}} 
\newcommand{\mponn}[1]{{\color{black}  #1}} 
\newcommand{\mpon}[1]{{\color{black}  #1}} 

\graphicspath{{./}{figures/}}

\begin{document}
\title{Cosmic-ray acceleration and escape from post-adiabatic Supernova remnants}

\author{R. Brose \inst{1,2}\fnmsep\thanks{Corresponding author, \email{robert.brose@desy.de}} \and
    M. Pohl  \inst{1,2}\and
    I. Sushch  \inst{1,3,4}\and
    O. Petruk  \inst{5,4,6}\and
    T. Kuzyo \inst{5} }

\institute{DESY, 15738 Zeuthen, Germany 
\and Institute of Physics and Astronomy, University of Potsdam, 14476 Potsdam, Germany
\and Centre for Space Research, North-West University, 2520 Potchefstroom, South Africa
\and Astronomical Observatory of Ivan Franko National University of Lviv, Kyryla i Methodia 8, 79005 Lviv, Ukraine
\and Institute for Applied Problems in Mechanics and Mathematics, Naukova 3-b, 79060 Lviv, Ukraine
\and Astronomical Observatory of the Jagiellonian University, Orla 171, 30-244 Kraḱow, Poland}
\date{Received ; accepted}



\abstract
{Supernova remnants are known to accelerate cosmic rays on account of their
non-thermal emission of radio waves, X-rays, and gamma rays. Although
there are many models for the acceleration of cosmic rays in Supernova remnants, the escape of cosmic rays from these sources is yet understudied.}
{We use our time-dependent acceleration code RATPaC to study the acceleration of cosmic rays and their escape in post-adiabatic Supernova remnants and calculate the subsequent gamma-ray emission from inverse-Compton scattering and Pion decay.}
{We performed spherically symmetric 1-D simulations in which we simultaneously solve the transport equations for cosmic rays, magnetic turbulence, and the hydrodynamical flow of the thermal plasma {in a volume large enough to keep all cosmic rays in the simulation.} The transport equations for cosmic-rays and magnetic turbulence are coupled via the cosmic-ray gradient and the spatial diffusion coefficient of the cosmic rays, while the cosmic-ray feedback onto the shock structure {can be} ignored. Our simulations span 100,000 years, thus covering the free-expansion, the Sedov-Taylor, and the beginning of the post-adiabatic phase of the remnant's evolution.}
{At later stages of the evolution cosmic rays over a wide range of energy can {reside outside of} the remnant, creating spectra that are softer than predicted by standard diffusive shock acceleration and feature breaks in the $10-100\,$GeV-range. {The total spectrum of cosmic rays released into the interstellar medium has a spectral index of {$s\approx 2.4$} above roughly $10$~GeV which is close to that required by Galactic propagation models}. We further find the gamma-ray luminosity to peak around an age of 4,000 years for inverse-Compton-dominated high-energy emission. Remnants expanding in low-density media emit generally more inverse-Compton radiation matching the fact that the brightest known supernova remnants - RCW86, Vela Jr, HESSJ1721-347 and RXJ1713.7-3946 - are all expanding in low density environments.}
 {}

\maketitle


\section{Introduction}
Supernova remnants (SNRs) are known to accelerate cosmic-rays (CRs) to relativistic energies \citep{Fermi.2013a}, and the highest energies are likely reached during the earliest phases of SNR evolution \citep{2013MNRAS.431..415B} and before the transition to the Sedov-phase \citep{2003A&A...403....1P}.

As soon as the peak maximum energy of a SNR is reached, the highest energetic CRs start to leak from the remnant. Whatever the CR spectrum inside the SNR at some point in time, the CR contribution of an SNR to the sea of Galactic CRs is given by the time integral of the CR leakage into the interstellar medium (ISM). Analytic calculations showed that the release spectra can be significantly softer than the spectra inside the SNRs \citep{2005A&A...429..755P,2010A&A...513A..17O}. Recently, \cite{2019arXiv190609454C} developed an analytic description for both the released CR spectrum and the spectrum of CRs remaining inside the remnant. They showed, that the spectrum inside the remnants can reassemble broken power-laws similar to the spectra observed in middle-aged remnants \citep{2009ApJ...698L.133A, Fermi.2013a}. At the same time, the release spectrum can be a $s=2$ power-law, if the acceleration spectrum has an index of $s<2$, or can have the same spectral index as the accelerated spectrum if $s>2$.

Models for the propagation of Galactic CRs indeed require injection spectra with a break at a few GV/c in rigidity that are hard at low energies and assume spectral indices around $s\approx 2.4$ above the break energy \citep{2009A&A...497..991P,2011ApJ...729..106T}, which are softer than those produced by linear diffusive shock acceleration (DSA). The electron injection spectra appear to be even softer than that above 30 GeV \citep{1998ApJ...493..694M} which most likely reflects electron energy losses prior to their release from the remnant \citep{2019arXiv190507414D}. Moreover, the gamma-ray emission of the middle-aged SNRs W44 and IC443 indicates CR spectra with an spectral index of $s\approx2.7$ at the highest energies \citep{2009ApJ...698L.133A, Fermi.2013a} and thus even softer than the injection spectra required by the propagation models. Further, most remnants seem to show spectral breaks in their high energy gamma-ray spectra with a break energy that is typically decreasing with increasing age \citep{2019ApJ...874...50Z}. 

Recent models for CR acceleration that include the CR feedback on the plasma flow, as well as the amplification of magnetic turbulence, are either steady-state calculations (e.g \cite{2014ApJ...789..137B}) or based on kinetic simulations that cover only very short time-scales \citep{Caprioli.2009}. The late phases of CR acceleration have been studied by \cite{2012APh....39...12Z} with a non-linear DSA model not considering the self-consistent amplification of turbulence. Reacceleration of CRs at slow shocks may arise  \citep{2016A&A...595A..58C}, as does a spectral modification due to an enhanced escape-rate of CRs \citep{Malkov.2011a}. \rb{Further, energy \mpon{transfer from} CRs to magnetic turbulence \citep{2019MNRAS.488.2466B} and deviations from spherical symmetry \citep{2019ApJ...881....2M} might lead to softer CR-spectra.}

We present a time-dependent test-particle calculation of CR acceleration over most of the lifetime of an SNR, including the beginning of the post-adiabatic phase. We show that the maximum energy of the accelerated CRs evolves only by one order of magnitude for Bohm-like diffusion but much more strongly than that, if the self-consistent amplification of Alfv\'en waves is taken into account. The spectra inside the remnant as well as the total production-spectra feature spectral breaks and softer spectral indexes as than those predicted by standard DSA. 

\section{Basic equations and assumptions}

\subsection{Cosmic rays}
We model the acceleration of cosmic rays using a kinetic approach in the test-particle approximation \citep{Telezhinsky.2012a,Telezhinsky.2012b,Telezhinsky.2013}, and we choose parameters for which the CR-pressure always stays below 10\% of the shock ram pressure. The time-dependent transport equation for the differential number density of cosmic rays $N$ \citep{Skilling.1975a} is given by:

\begin{align}
\frac{\partial N}{\partial t} =& \nabla(D_r\nabla N-\mathbf{u} N)\nonumber\\
 &-\frac{\partial}{\partial p}\left( (N\dot{p})-\frac{\nabla \cdot \mathbf{ u}}{3}Np\right)+Q
\label{CRTE}\text{ , }
\end{align}
where $D_r$ denotes the spatial diffusion coefficient, $\textbf{u}$ the advective velocity, $\dot{p}$ energy losses and $Q$ the source of thermal particles.

We solve this transport equation in a frame co-moving with the shock. The radial coordinate is transformed according to $(x-1)=(x^*-1)^3$, where $x=r/R_{sh}$. For a equidistant binning of $x^*$ this transformation guarantees a very fine resolution close to the shock and an outer grid-boundary that extends to several tens of shock-radii upstream for $x^*>>1$. Thus all accelerated particles can be kept in the simulation domain. 

The background of Galactic cosmic rays is introduced as initial condition outside of the remnant and as boundary condition for the differential cosmic-ray density very far upstream, which is equivalent to assuming an infinite supply of cosmic rays at the boundary.
We describe the spectrum of hadronic CRs as a power law in total energy, modified at low energy by the particle speed, $\beta$. The electron spectrum is a log-parabola at low energies, 
\begin{align}
    N_\mathrm{PR}(E) &= N_1\beta(E) (E+mc^2)^{s_1}\label{Eq:BackgroundPRs}\\
    N_\mathrm{EL}(E) &= 
        \begin{cases}
            \frac{N_2}{E}\exp\left(-\frac{\log^2({E}/{E_c})}{\sigma}\right)\text{ for }E\leq E_B\\
            N_3E^{s_2} \text{ for }E>E_B
        \end{cases}\label{Eq:BackgroundELs}
\end{align}
Both electron and proton background spectra can be directly measured above a few GeV, where solar modulation is unimportant. Whereas the galactic electron spectrum can be constrained by measurements of diffuse radio emission, the spectral slope of the proton spectrum at low energies remains unclear. At high energies the local CR spectrum has an index $s_1 = -2.75$ which gives a spectrum harder than $E^{-2}$ at low energies; we chose the normalization in accordance with \cite{2002ApJ...565..280M}. 
We also investigated an alternative, softer background spectrum and discuss the differences in appendix \ref{sec:AlternateBackground}.

For the electron spectrum, we fitted the spectra given in \cite{2011MNRAS.416.1152J} for a galactocentric radius of $6.5\,$kpc with expression (\ref{Eq:BackgroundELs}) as the spectrum shows a continuous change in the index below $4\,$GeV. The spectral index at high energies is found to be $s_2 = -3.04$ and $E_B=5\,$GeV. \rbn{These spectra are compatible with direct observations of the electron spectra in the local ISM by Voyager 1, which also show spectra harder that $s=2.0$ at low energies \citep{2016ApJ...831...18C}.}

\subsubsection{Injection} 
We inject particles according to the thermal leakage injection model \rb{\citep{Blasi.2005a,1998PhRvE..58.4911M}}. Here the efficiency of injection $\eta_i$ is given by
\begin{align}
\eta_i = \frac{4}{3\sqrt{\pi}}(\sigma-1)\psi^3e^{-\psi^2}\text{ , }
\end{align}
where $\sigma$ is the shock compression ratio and $\psi$ is the multiple of the thermal momentum we inject particles at. \rb{\citet{2019ApJ...872..108H} suggests an additional dependence of $\eta_i$ on the shock Mach number. However, \mpon{this behaviour} is observed only for low-Mach-number shocks and will be ignored here.} \citet{2013MNRAS.430.2873R} used a spherical-harmonics expansion of the cosmic-ray Fokker-Planck equation to find a quasi-universal behaviour of shocks irrespective of the magnetic-field orientation very far upstream of the shocks, which suggests that injection is only weakly dependent on the shock orientation\footnote{\rb{\mpon{It is conjecture only that} the morphology of the non-thermal emission from SN 1006 is often attributed to the local magnetic-field angle \citep{2003A&A...409..563V}.}}. Recent MHD-PIC simulations seem to support this notion \citep{2018MNRAS.473.3394V}, and so we do not differentiate between quasi-parallel and quasi-perpendicular shocks. This injection scenario should not be taken literally, in particular not for electrons, for which pre-acceleration to a few tens of MeV is required and established at the shock \citep{Matsumoto2017,2018PhPl...25h2103L,2019ApJ...878....5B}. We are interested in particles at energies well above $100$~MeV, and so the particulars of that pre-acceleration can be ignored.

To be noted is that $\eta_i$ is not the only factor controlling injection. The shock velocity and the upstream density determine the rate with which particles pass through the shock. As the shock speed decreases, the injection rate decreases as well. Whereas in the wind-zone of a core-collapse supernova a large part of the particles is injected early in the evolution, for a type-Ia SNR expanding into a uniform ISM, the area-integrated injection rate increases with time on account of the enlarging shock surface. As a result the most recently injected particles dominate the volume-averaged particle spectra.

Usually $\eta_i$ is assumed to be constant, but it is not inconceivable that $\eta_i$ might change with time, as it is a simple parametrization of nonlinear microphysical processes operating at the shock front \citep{2003A&A...409..563V,2016MNRAS.462.3104P}. To investigate the effects of a time-dependent injection efficiency, we also consider a variable $\eta_i$ of the form
\begin{align}
    \eta_{i,t}(t) = \eta_i\left(\frac{t}{t_0}\right)^{a} \text{ . }\label{eq:InjOverTime}   
\end{align}

%

\subsection{Magnetic field and diffusion coefficients}\label{sec:MF}
To obtain the large-scale magnetic field, we assume it is dynamically irrelevant and hence solve the induction equation following \citep{Telezhinsky.2013}. The magnetic field is assumed to be constant in the upstream of the shock at a value of $5\,\mu$G. The field strength in the immediate downstream of the shock is $\sqrt{11}\cdot5\,\mu$G$\approx16\,\mu$G\footnote{We consider a magnetic field with equally strong parallel and perpendicular components. The parallel direction is not compressed at the shock, hence the compression-factor is $\sqrt{11}$ instead of $4$.}. 

Observations indicate an amplification of the magnetic field in the downstream to several $100\,\mu$G for very young SNRs \citep{2003A&A...412L..11B}, at least part of which likely arises from MHD processes at and behind the shock \citep{2007ApJ...663L..41G}. Theory suggests efficient magnetic-field amplification also in the precursor of the remnant \citep{2000MNRAS.314...65L, 2001MNRAS.321..433B}. However, the nonlinearity imposed by this magnetic-field amplification does not allow simple scaling (or averaging) of the results to the entire population of Galactic cosmic rays; modeling it hence must be beyond the scope of this paper, and we checked magnetic turbulence to only reach levels of $\delta B \approx B_0$. \rb{\mpon{The amplification of magnetic field well beyond its initial value imposes energy loss of CRs that} can steepen the spectrum of the accelerated CRs. \mpon{\citet{2019MNRAS.488.2466B} find a spectral steepening that scales with $v_\mathrm{shock}/c$ and is hence relevant only during the early phases of SNR evolution.} Here, this effect can be neglected as the amplification factor of the field reaches only a value of $\approx1$, \mpon{considerably less than assumed by \citet{2019MNRAS.488.2466B}}.}

We choose two models for the diffusion that can be considered as limiting cases. In one case we apply Bohm-like diffusion at the shock and have a exponential transition to the Galactic diffusion coefficient further upstream at $2R_{sh}$ (following \citet{Telezhinsky.2012b}). 

In the second case we solve in parallel a transport equation for the magnetic turbulence spectrum, assuming Alfv\'en waves only, and thus calculate the diffusion coefficient self-consistently \citep{2016A&A...593A..20B}. In this case the diffusion coefficient is time-dependent. However, we inject only few particles and thus the magnetic field is not amplified above the background magnetic field value in this case. The former case can be considered as a lower limit, and the latter ansatz as an upper limit for the diffusion coefficient. 

\rb{In both cases, the diffusion process is assumed to be isotropic. In reality $D_\perp << D_\parallel$, and so the escape of particles along the field lines is enhanced. This situation can be neglected as long as the coherence length of the magnetic field is smaller than the size of the remnant \mpon{\citep{2018MNRAS.479.4526L}.} 
In this study we are mainly interested in middle-age and old remnants whose size is sufficient to justify the isotropy assumption.}

\subsection{Hydrodynamics}
The evolution of an SNR without CR-feedback can be described with the standard gas-dynamical equations
\begin{align}
\frac{\partial }{\partial t}\left( \begin{array}{c}
                                    \rho\\
				    \textbf{m}\\
				    E
                                   \end{array}
 \right) + \nabla\left( \begin{array}{c}
                   \rho\textbf{v}\\
		   \textbf{mv} + P\textbf{I}\\
		   (E+p)\textbf{v} 
                  \end{array}
 \right)^T &= \left(\begin{array}{c}
                    0\\
		    0\\
		    L
                   \end{array}
 \right)\\
 \frac{\rho\textbf{v}^2}{2}+\frac{P}{\gamma-1}  &= E \text{,}
\end{align}
where $\rho$ is the density of the thermal gas, $\textbf{v}$ the plasma velocity, $\textbf{m}=\textbf{v}\rho$ the momentum density, $P$ the thermal pressure of the gas, $L$ the energy losses due to cooling, and $E$ the total energy density of the ideal gas with $\gamma=5/3$. This system of equations is solved under the assumption of spherical symmetry in 1-D using the PLUTO code \citep{2007ApJS..170..228M}. The non-equilibrium cooling function, $L$, is taken from \cite{1993ApJS...88..253S}.

In this work we display results for type-Ia supernova explosions and discuss the basic differences to core-collapse explosions in appendix \ref{sec:CoreCollapse}. Therefore we initiate the simulations with ejecta profiles
\begin{align}
 \rho_\mathrm{SN} =& A\exp(-v/v_e)t_i^{-3}\text{ and } v = r/t_i \\
 \text{ with } & v_e = \left( \frac{E_\mathrm{ex}}{6M_\mathrm{ej}} \right)^{1/2} \text{ and } A =\frac{6^{3/2}}{8\pi}\frac{M_\mathrm{ej}^{5/2}}{E_\mathrm{ex}^{3/2}}
\end{align}
as initial conditions \citep{1998ApJ...497..807D}. Here $t_i=20\,\mathrm{yrs}$ is the start time of our simulation, $M_\mathrm{ej}=1.4M_\mathrm{sol}$ the ejecta mass, $E_\mathrm{ex}=10^{51}\,$erg the explosion energy, and $r$ the spatial coordinate.

The initial age of about 20 years is rather large but the solution quickly converges against solutions with a lower initial age. In any case, we are mainly interested in the later stages of the evolution. The density of the ambient medium was chosen to be $0.4\,\text{cm}^{-3}$ unless stated differently.

The results of the hydro simulation for the density, velocity, pressure and temperature distributions are then mapped onto the spatial coordinate of the CR and turbulence grid respectively. The shock, that is typically a few bins wide in the hydro-solution, needs to be resharpened in order o guarantee a realistically high acceleration rate from GeV to TeV energies. This procedure is repeated for each time step of the CR and turbulence grid. One of these time steps typically requires many time steps of the hydro-solver.




\section{Results}
\begin{figure}[h]
\includegraphics[width=0.5\textwidth]{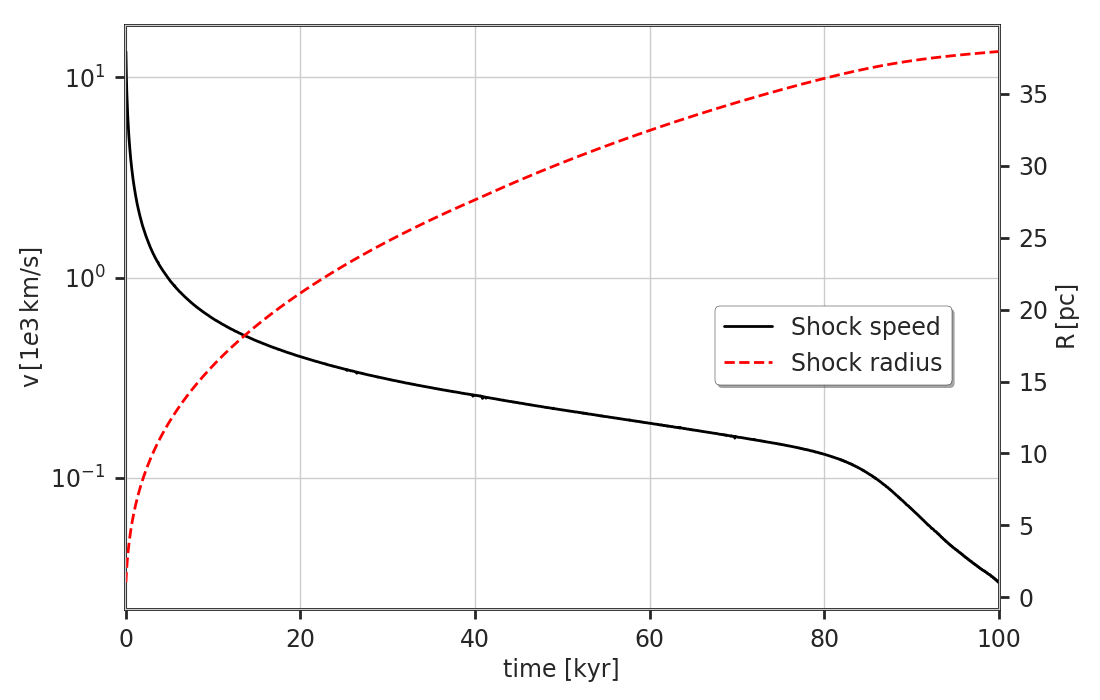}
\caption{The curve shows the radial position of the forward shock as function of time. The transitions between free-expansion and the Sedov phase, and between Sedov and the post-adiabatic phase, appear roughly at 1.3\,kyrs and 35\,kyrs, respectively.}
\label{fig:EvoSh}
\end{figure}

Using the framework described above, we followed the evolution of the remnant for $100,000\,$years. The shock speed was $9200\,$km/s after $100\,$yrs. The transition to the Sedov phase happened after $1,300\,$years when the swept-up mass was approximately $10\, M_\mathrm{sol}$ and the shock speed $v_\mathrm{ shock}=2350\,$km/s. The remnant reached the post-adiabatic phase after $35,000\,$years\footnote{This is the time $t_\mathrm{*}$ when radiative losses of a fluid element during $t_\mathrm{*}$ are comparable to its initial thermal energy \citep{1972ApJ...178..159C}, so the shock cannot be considered adiabatic any more.}  with $v_\mathrm{shock}=300\,$km/s. After roughly $85,000\,$years ($v_\mathrm{shock}=130\,$km/s) the shock compression ratio started to fall significantly below $4$, and at the end of the simulation it had a value of $3.15$. Fig.~\ref{fig:EvoSh} displays the time evolution of the speed and position of the forward shock (cf. Fig.~4a by \citet{2016MNRAS.456.2343P}). 

Already after $10,000\,$years the shock speed is down to $640\,$km/s, and so the difference in speed between the end of the Sedov phase and the post-adiabatic phase is only a factor of ten. The maximum energy, that can be reached during the post-adiabatic phase, should then be only one order of magnitude lower than that during the Sedov phase.

In this section we will first present results for a post-adiabatic remnant under different assumptions about CR diffusion. We shall describe the escape of CRs from the remnant, the reacceleration of pre-existing CRs, and the inverse Compton and Pion-decay gamma-ray spectra.

\subsection{Escape}
To distinguish the escape of CRs produced in the remnant and reacceleration of Galactic CRs, in this section we set the density of Galactic CRs to zero.

We first evaluate the evolution of the CR-spectra for the two diffusion scenarios that we introduced in subsection~\ref{sec:MF} -- Bohm-like diffusion in a $5$-$\mu$G field around the remnant and a diffusion coefficient obtained from the self-consistent amplification of Alfv\'enic turbulence. 

\begin{figure}[h]
\includegraphics[width=0.5\textwidth]{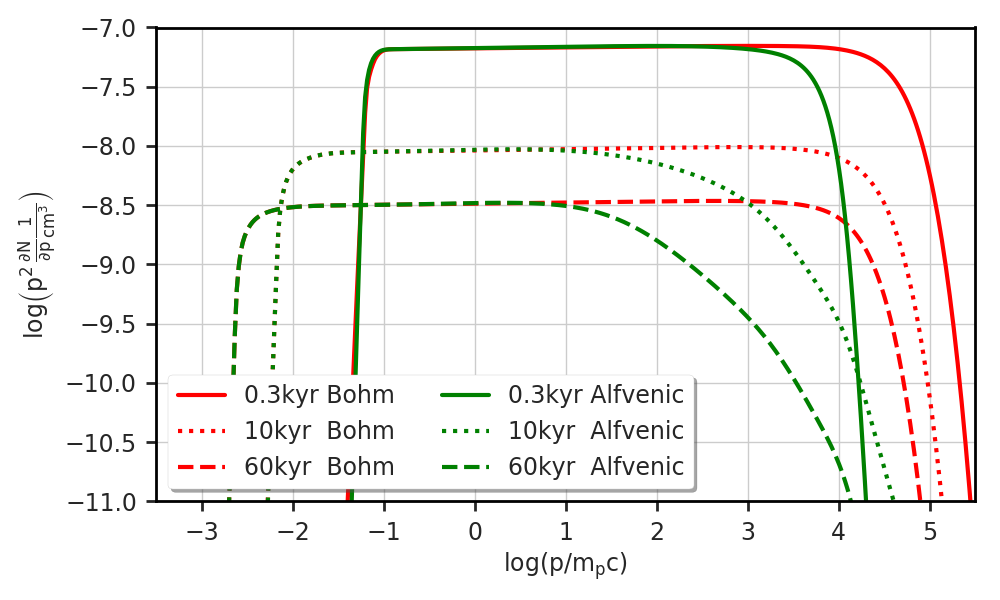}
\caption{Volume-{averaged} downstream proton spectra for Bohm-like (red) and self-consistent (green) diffusion. The background of Galactic CRs is neglected.} 
\label{fig:BohmSpecDown}
\end{figure}

Figure \ref{fig:BohmSpecDown} shows the evolution of the volume-{averaged} cosmic-ray spectrum in the downstream region at three points in time that represent the three stages of SNR evolution. These downstream spectra represent the particle population that is mainly responsible for the detectable emission from the remnant.

To be noted from the figure is that for Bohm-like diffusion (red lines) the spectra are power laws with the typical test-particle index of $s=2.0$ at all stages of the evolution. The normalization decreases as the remnant decelerates and particles get injected at lower energies, where most particles reside. The maximum energy decreases approximately from $20\,$TeV to $5\,$TeV. The decrease in maximum energy is small, because for a constant upstream magnetic-field strength the maximum energy decreases very slowly in the Sedov phase; in fact the shock speed drops only by a factor 20 between the beginning of the Sedov phase and the sharp decrease of the shock-velocity at an age of $85,000\,$yrs, and the time available for particle acceleration at the late, slow shock is roughly one order of magnitude longer than the lifetime of the fast shock.

With explicit treatment of turbulence transport and Alfv\'enic diffusion, the decreasing normalization of the CR density and hence the CR pressure gradient reduces the driving of turbulence. As a consequence, the diffusion coefficient and the timescale for acceleration increase, and the maximum energy falls from $5\,$TeV to $10\,$GeV. As it takes time for particles to escape from the remnant, high-energy particles are still present in small numbers, and the cosmic-ray spectra display a break at the momentary maximum energy and are soft at higher energies. Figure \ref{fig:EscapeSpecTotNorm} shows the total number of CRs for the Alfv\'enic scenario at different times. It can be seen that all CRs with an energy above $1\,$TeV are produced within the first $10\,$kyrs of the remnants evolution. At the later stages CRs from the downstream are escaping to the upstream which forms softer spectra inside the remnant. The break in the spectrum occurs at the energy the SNR is currently able to accelerate CRs to. This behaviour was also obtained in the analytic calculations of \cite{2019arXiv190609454C}, who found that high-energy particles that already escaped the acceleration process can be trapped inside and close to the remnant and introduce a spectral break in the spectrum of particles present inside the SNR. The particle spectrum above the break energy is not a simple power-law but can be reasonably well approximated with a $s=2.7$ power-law index. Interestingly, this spectral index is in rough agreement with that measured in IC443 and W44 \citep{Fermi.2013a}. In both remnants, the CR spectra also feature a spectral break around a few hundred GeV. We acknowledge that both IC443 and W44 interact with dense material and locally have a wide range of evolutionary age. Our model assumes spherical symmetry and an external medium with a constant density. In first-order approximation, a composite model based on spectra calculated for different ages should permit a rough comparison with interacting SNRs though.

Our findings are compatible with analytic calculations by \cite{2012PhPl...19h2901M}. There, the authors suggested that an evanescence of Alfv\'en waves due to strong neutral-charged collisions in SNRs interacting with dense environments leads to a spectral steepening by exactly one power ($\Delta s = 1.0$). In this case particles above a break energy around $10\,$GeV can escape the shock precursor, and the spectra above the break energy feature softer spectra. The authors noted that their model well explains the gamma-ray observations of W44 \citep{Fermi.2013a,2011ApJ...742L..30G} but overpredicts the spectral steepening observed in IC443 and W28. Here the spectral index changes only by $\Delta s = 0.6-0.7$ \citep{Fermi.2013a,2009ApJ...698L.133A,2010ApJ...718..348A} in agreement with our findings.

In our model the reason for the softening of our spectra is also an evanescence of Alfv\'en waves but caused by reduced driving of turbulence at later stages of the remnants evolution, which introduces additional nonlinearity and time-dependence. A second difference to the scenario of \citet{2012PhPl...19h2901M} is that escaping particles are still scattered and confined to the shock region in our simulations.

\begin{figure}[h]
\includegraphics[width=0.48\textwidth]{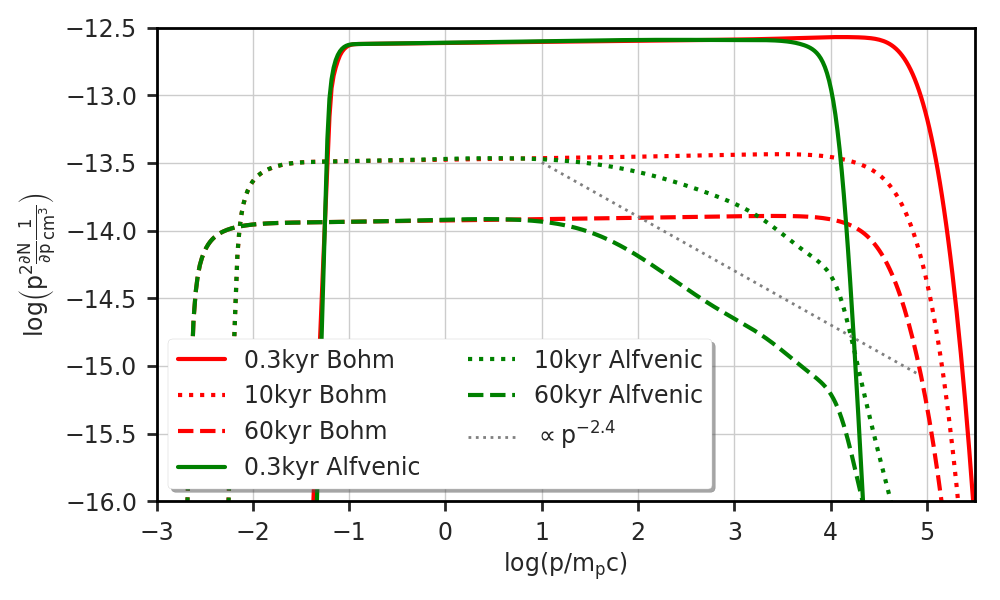}
\caption{The total volume-{averaged} proton spectrum in the case of Bohm-like (red) and self-consistent (green) diffusion. Galactic CRs are ignored.}
\label{fig:EscapeSpecTot}
\end{figure}

\begin{figure}[h]
\includegraphics[width=0.48\textwidth]{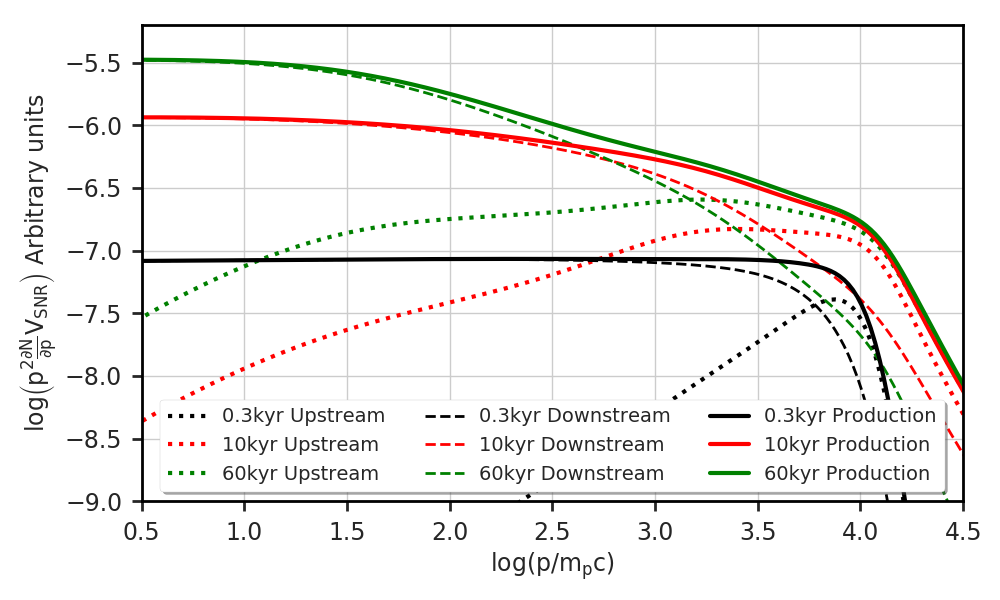}
\caption{The proton number-spectra for the Alfv\'enic diffusion scenario at different times. Solid lines represent the total, dashed the downstream and dotted the upstream spectra.}
\label{fig:EscapeSpecTotNorm}
\end{figure}

We can also evaluate the total production spectrum of CRs, including those that left the remnant. Our spatial grid extends to $65\, R_\mathrm{shock}$, and all accelerated particles stay in the simulation domain. Then, the CR spectrum integrated over the whole simulation domain represents the total CR yield of the SNR. These total production spectra then represent the spectrum of CRs that the remnant injects into our galaxy\footnote{If all acceleration would stop at this given time so the particle spectrum would not be modified any more}. Still, spectra above the current maximum energy are only weakly modified during later times of the SNRs evolution. where they are subject to further modification during galactic propagation. These spectra are typically different from the downstream spectra that represent the particle population responsible for the emission detected from the remnant. Figure \ref{fig:EscapeSpecTot} shows the total production spectra for the two diffusion regimes discussed in this section. Again, for Bohm-like diffusion the spectrum is well represented by a power-law with index $s=2.0$. In the Alfv\'enic regime, a spectral break occurs at $\approx10\,$GeV after $50,000\,$yrs, where the spectral index changes approximately from $s=2.0$ to $s=2.4$.

The total proton spectrum coincides well with that deduced from fits of cosmic-ray data with Galactic propagation models, where the required injection spectrum tends to be slightly harder than $E^{-2}$ below a few GeV, transitioning to a softer index $s\simeq 2.4$ above 10~GeV \citep{2009A&A...497..991P,2011ApJ...729..106T}. Our results satisfy that expectation for the entire energy range above $10$~GeV where direct measurements are possible and solar modulation is weak or absent, even though we conduct test-particle calculations for which the spectra of CRs at the shocks are always standard DSA $s=2.0$ spectra. They also provide a natural explanation for the break in the cosmic-ray source spectrum that was introduced ad-hoc in the fits of the cosmic-ray data and appears here on account of the explicit treatment of turbulence driving upstream of the shock. All changes in the spectral index arise from the time evolution of SNRs, the turbulence spectra, and consequently the maximum energy of accelerated particles. Simple scaling relations such as the assumption of Bohm diffusion significantly modify the model expectations, as do steady-state descriptions of the turbulence level and the diffusion coefficient. The spectral indices obtained for Alfv\'enic diffusion depend on the particulars of wave growth, damping, and cascading. They reflect the balance between the reduction rate of the maximum energy of CRs at the shock and the escape rate of high-energy CRs from the SNR \citep{2010A&A...513A..17O}.

\rb{Similar soft spectra \mpon{as those due to the decay of turbulence} can be obtained, if the maximum \mpon{particle energy significantly varies along} the shock surface. \cite{2019ApJ...881....2M} discuss \mpon{the expansion of a SNR in} a homogeneous magnetic field, \mpon{assuming that efficient acceleration only takes place at} two polar caps, as the shock normal has to be sufficiently parallel to the magnetic field for DSA to operate. The spectra observed along some line-of-sight through the acceleration region will then sample regions with different maximum energies and be softer than those predicted by standard DSA. Both methods rely on \mpon{a strong variation of} the maximum energy, either spatially or temporally, and the softer spectra arise from superposition of spectra \mpon{at different locations or times}.}

Our total production index above the break energy differs from the results obtained by \cite{2019arXiv190609454C}. Their model predicts a spectral index of $s=2$ for the total production spectrum as the spectrum we obtain through acceleration at the shock is always a $s=2$ power-law. However, they assumed that a constant fraction of the SNRs kinetic energy is transferred into CRs. As a result, the fraction of thermal particles injected as CRs is not a constant in their calculation and depends on the shock ram-pressure and the maximum particle energy\footnote{Their spectral normalization is $\propto \xi_{CR} \rho_0 v_{sh}^2$}. The effect of an increasing shock surface - meaning more particles injected into the acceleration process at later times - is canceled by the impact on the proton spectral normalization of the decreasing shock speed, leading to $s=2$. In our case the fraction of thermal particles injected as CRs is a constant which leads to a different time-dependence of the spectral normalization. Thus, our resulting spectra can be softer than $s=2$ even if our acceleration spectrum has the standard DSA spectral index.

\begin{figure}[h]
\includegraphics[width=0.48\textwidth]{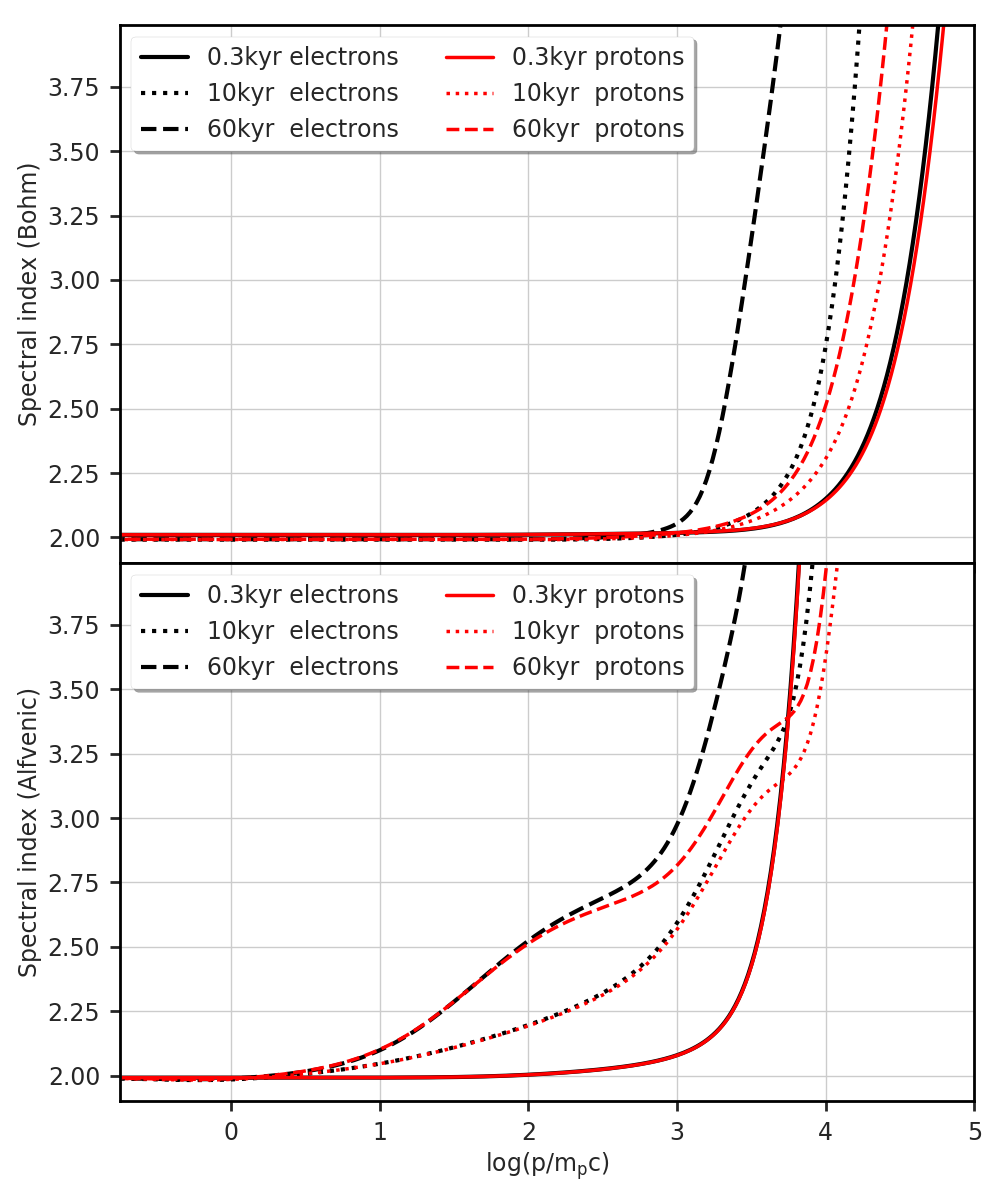}
\caption{Running power-law index of the electron (black) and proton (red) spectrum as function of energy displayed at different ages of the remnant. The Bohm-like diffusion scenario is shown in the top-panel and the Alfv\'enic-turbulence scenario in the bottom-panel.}
\label{fig:Electrons}
\end{figure}

Figure \ref{fig:Electrons} shows the distribution of the spectral index of the CR electrons compared to that of the protons. To be noted from the figure is that beyond an age of about $300\,$yrs the electron spectra are softer than the proton spectra on account of synchrotron cooling, which is in agreement with the findings of \cite{2019ICRC...36...59D}. However, the difference in spectral index is a strong function of the turbulence model, in particular its time dependence, and cannot be described as a simple shift in the spectral index. We emphasize again, that the decay of turbulence and the subsequent escape of particles is essential for the formation of soft particle spectra and spectral breaks, even if the acceleration mechanism at the shock produces a standard DSA spectrum. Steady-state models and simple scaling relations miss part of these features.

\section{Reacceleration}
It has been proposed by several authors \citep[e.g.][]{2015ApJ...800..103T,2016A&A...595A..58C, 2018JPlPh..84c7101C} that part of the particles accelerated at an SNR might be reaccelerated, pre-existing CRs. They may even dominate the gamma-ray output of W44 and IC443 \citep{Fermi.2013a}.

To test this scenario we included background CRs into our simulations. They should become relevant when the injection rate of particles into DSA at the shock is very low. To numerically control this threshold effect, we artificially varied the injection efficiency over time according to Eq. (\ref{eq:InjOverTime}). We adopted $a=-1.0$ for the power-law dependence of the injection with time, $\eta_i\propto t^a$, and took 20 years as our reference point. After $2500\,$ years we inject with an efficiency that is $0.8\,$\% of that used initially. From this point in time, the normalization of the CRs injected at the shock is low enough that background CRs can dominate the acceleration and emission. The level of background CRs that we have in our simulation domain was negligible for the results presented in the previous section. It has been argued that the level of background CRs might be increased inside the molecular clouds that some remnants are interacting with. However, the possible level of background CRs inside the clouds is constrained by observations of the gamma-ray emission from these clouds and found to be comparable to that measured outside of the clouds
\citep{2014A&A...566A.142Y}.

\begin{figure}[h]
\includegraphics[width=0.5\textwidth]{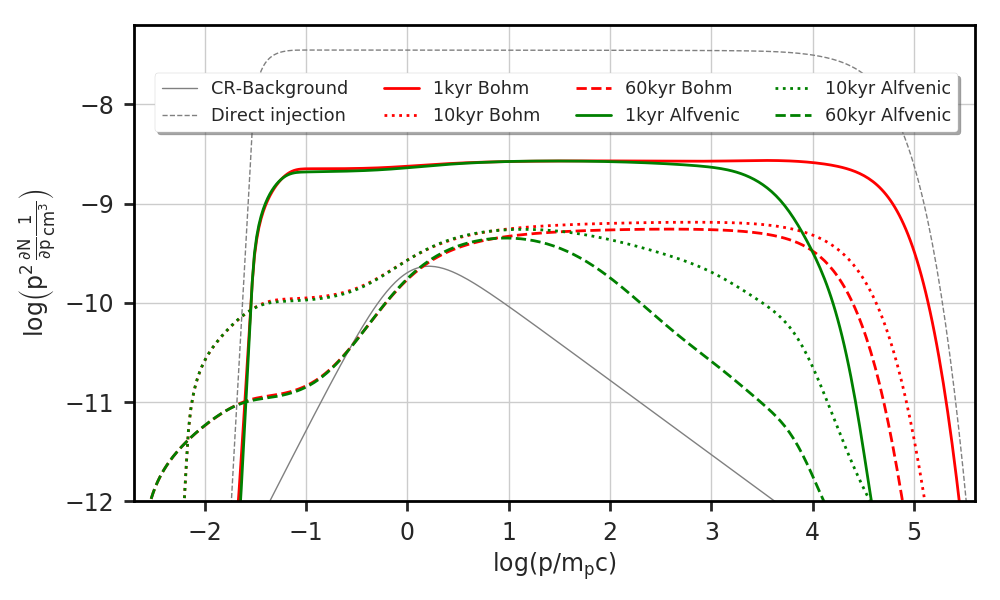}
\caption{Volume-{averaged} downstream proton spectra for Bohm-like (red) and self-consistent (green) diffusion at different times. We include background CRs and artificially decrease with time} the injection fraction of thermal particles at the shock. The CR-density at $300\,$yrs for non-decreasing injection (direct) is shown as a dotted-grey line for comparison.
\label{fig:BohmB}
\end{figure}

Again, we investigate the two regimes of Bohm-like and self-generated Alfv\'enic diffusion. All other simulation parameters are kept the same as in the previous section. Figure \ref{fig:BohmB} illustrates the CR spectra inside the remnant for the two diffusion models for the three phases of the SNRs evolution. 

Again, for Bohm-like diffusion the spectrum of accelerated CRs remains a $s=2.0$ power-law with only a modest change in the maximum energy. After $2500\,$yrs the background CRs fill the entire simulation volume, and the normalization of the CR-spectrum does not decrease further as it did without background CRs (cf. Fig. \ref{fig:BohmSpecDown}). Additionally, the low-energy part of the spectrum is dominated by freshly injected CRs as our imposed background CR-spectrum is very hard below $1\,$GeV (see Eq. \ref{Eq:BackgroundPRs}), and freshly injected CRs dominate the spectrum up to $\approx100\,$MeV. However, the shape of the low energy part of the spectrum strongly depends on the parametrization of the CR spectrum below $1\,$GeV which is not directly measurable due to the effects of solar modulation.

The situation looks similar for self-generated Alfv\'enic turbulence. The main difference is, that the accelerated background CRs provide too little turbulence to be contained by the remnant. A higher level of background CRs would enhance the amplification of turbulence and hence the confinement of CRs but a much higher background flux is in contradiction to direct observations \citep{2014A&A...566A.142Y}. The particle spectrum extends to about $100\,$GeV before it cuts off with a spectral index $s>-2.7$. This break energy coincides with that deduced for W44 and IC443 \citep{Fermi.2013a}. 

\begin{figure}[h]
\includegraphics[width=0.5\textwidth]{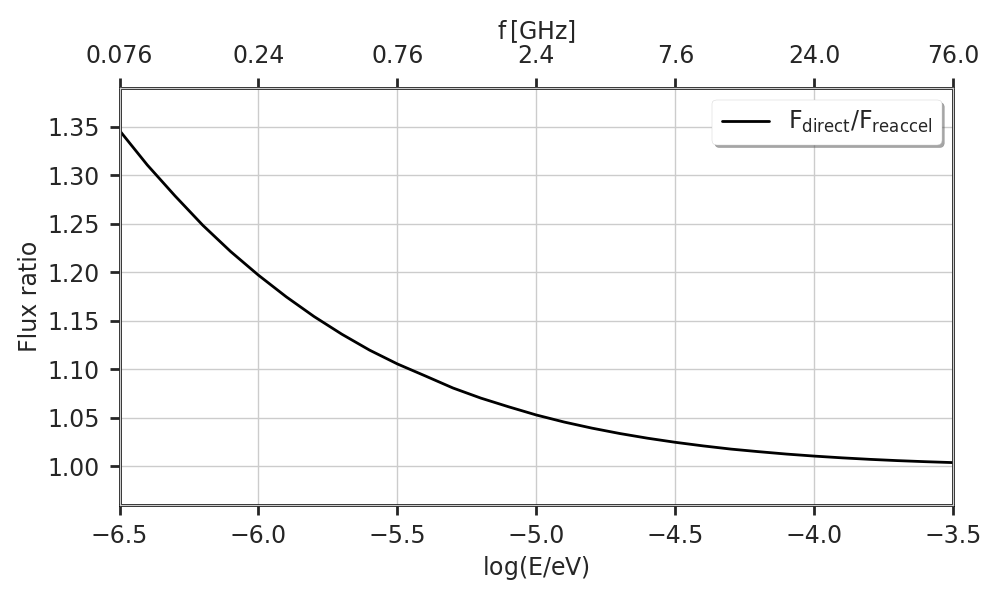}
\caption{The ratio of the radio flux produced by directly accelerated and reaccelerated electrons, normalized above $75\,$GHz. The assumed magnetic-field strength in the downstream region is $16\,\mu$G.}
\label{fig:Radio}
\end{figure}

In fact, both reacceleration of CRs from the sea of background CRs and the escape of CRs from radiative remnants provide similar signatures at higher energies. Both produce a spectral break at energies between $10-100\,$GeV with soft spectra at higher energies, with a slightly stronger turnover in the case of reacceleration. The emission signature will differ mostly at radio energies where either freshly injected CRs or background CRs dominate, which yield different spectra as shown in Figure \ref{fig:Radio}.

If freshly accelerated electrons dominate, the radio spectrum will be a featureless power law with index $\alpha=0.5$. The situation is different for reaccelerated electrons as the diffuse Galactic radio emission indicates electron spectra harder than $s=2.0$ below $4\,$GeV. This hard spectrum is retained when electrons get reaccelerated whereas the parts of the spectrum that are softer than $s=2.0$ change to $s=2.0$ during the reacceleration process. \mponn{If reaccelerated electrons are supposed to be more important than freshly accelerated electrons above a few GeV, then we would observe freshly accelerated electrons at very low energies with a $p^{-2}$ spectrum, until reaccelerated electrons take over with a harder spectrum reflecting that of electrons in the ISM, until around a few GeV the electron spectrum turns over to $p^{-2}$ again. The latter transition would appear in the radio synchrotron spectra} right in the observable frequency band between $100\,$MHz and $50\,$GHz, depending on the strength of the magnetic field. 
We note that remnants that are discussed to reaccelerate CRs, like W44, show remarkably featureless power-law radio spectra \citep{2007A&A...471..537C}.

\rb{\cite{2019MNRAS.489..108C} recently argued that the gamma-ray emission from the SNRs SN1006, Vela Jr. and RX J1713.7-3946 might be explained in terms of IC-emission from reaccelerated galactic electrons without the need of a contribution from freshly accelerated electrons. \rbn{Their ansatz and assumptions for the electron background spectrum are similar to \mponn{those} used in this paper but they did not investigate the radio emission that would arise from the reacceleration of galactic electrons.} As the magnetic fields in this three remnants are 
\mponn{likely between $10\,\mu$G and $100\,\mu$G and hence} comparable to the $16\,\mu$G we assumed in figure \ref{fig:Radio} \citep{2018A&A...612A...6H, 2018A&A...618A.155S, 2003A&A...412L..11B}, the radio spectra of these remnants should feature \mponn{turnovers described above. While the radio data for Vela Jr. and RX J1713.7-3946 are too sparse to draw any conclusion, it is clear that there is no evidence for a softening in the radio spectrum of SN1006; if anything, there might be a hardening with increasing energy \citep{2008ApJ...683..773A}}.}

\begin{figure}[h]
\includegraphics[width=0.5\textwidth]{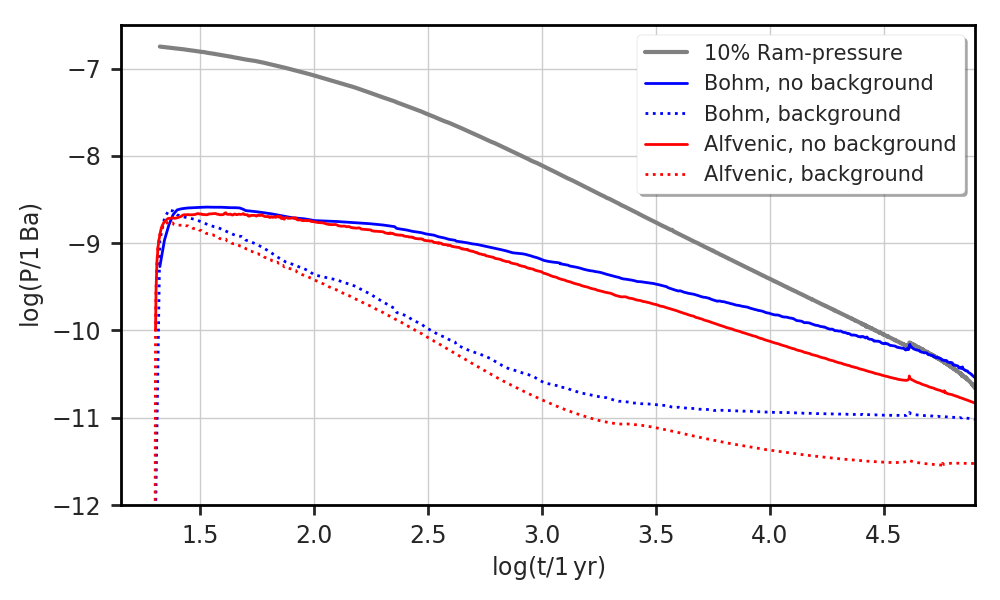}
\caption{Evolution of the CR pressure over time for the two diffusion scenarios and with/without the Galactic CR background. The grey line indicates 10\% of the shock ram pressure.}
\label{fig:PressureEvo}
\end{figure}

Figure \ref{fig:PressureEvo} shows the CR pressure in our simulations. The CR pressure stays well below 10\% of the shock ram pressure (grey line), and so our test-particle assumption is valid most of the time. Only for Bohm-like diffusion CRs become dynamically important at the end of the evolution, when the SNR enters the dissipative phase. It is likely that at this point also magnetic field is dynamically relevant \citep{2016MNRAS.456.2343P}.

Background CRs change the time-evolution of the CR-pressure which starts to fall less rapidly as soon as the re-accelerated background CRs begin to dominate the particle distribution at the shock. The initial rapid decline of the CR pressure in simulations with background CRs arises from our choice of a decreasing injection efficiency, $\eta_i \propto t^{-1}$, in these simulations. In any case, CRs start to be dynamically important at very late time as also suggested by \cite{2016ApJ...827L..29S}

\section{Illuminated clouds}

\begin{figure}[h]
\includegraphics[width=0.5\textwidth]{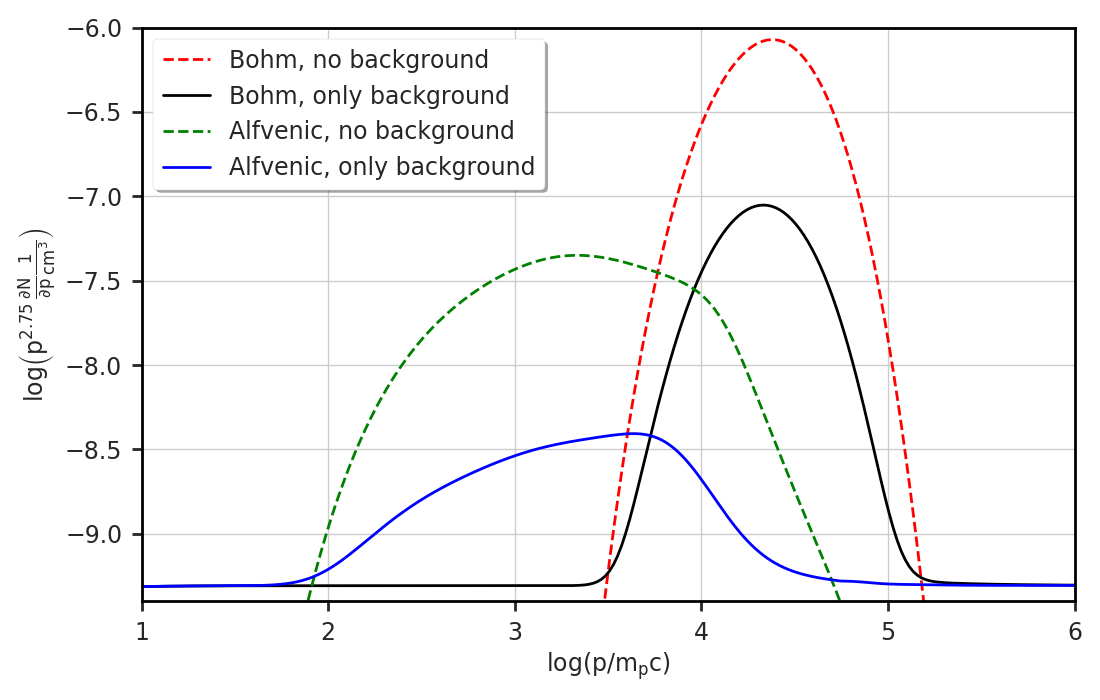}
\caption{CR spectra about $7.5\,$pc ahead of the forward shock at an age of $37\,$kyrs. Red (black) lines refer to Bohm-like diffusion without (with) background CRs, and green (blue) lines stand for Alfv\'enic diffusion without (with) background CRs. Note that the y-axis is scaled with $p^{2.75}$.}
\label{fig:Escape}
\end{figure}

Escaping cosmic rays do not only replenish CRs in the Galaxy, but may also illuminate molecular clouds close to their mother SNRs, leading to intense gamma-ray emission \citep{2007ApJ...665L.131G}. This illumination effect is a clear indicator of the acceleration of hadrons and has been observed in W28 \citep{2018ApJ...860...69C}. The spectra of escaping CRs are typically different from those of CRs inside the remnant. Figure \ref{fig:Escape} illustrates the spectra of escaped CRs $7.5\,$pc from the SNR shock in a $2\,$pc thick region at an age of $37\,$kyrs for the four setups (Bohm-like/Alfv\'enic diffusion, acceleration/reacceleration) discussed before. 

The spectra for Bohm-like diffusion resemble narrow bumps that can be described as log-parabolas.  
If the maximum energy of accelerated particles evolves quickly, as it does for diffusion in the time-dependent Alfv\'enic turbulence, the spectra are much wider and have a lower amplitude. 
Over a wide range the far upstream spectrum of reaccelerated CRs is harder than $s=2.75$ - which corresponds to a horizontal line in figure \ref{fig:Escape} - with a cut-off near 10~TeV. Freshly accelerated CRs have a higher flux and a roughly symmetric spectrum. 
Apart from the normalization and the slightly different spectral index there is no spectral signature that would clearly distinguish reaccelerated CRs from particles injected from the thermal pool and accelerated at the shock.  

\section{Gamma-ray emission}

\begin{figure}[ht]
\includegraphics[width=0.5\textwidth]{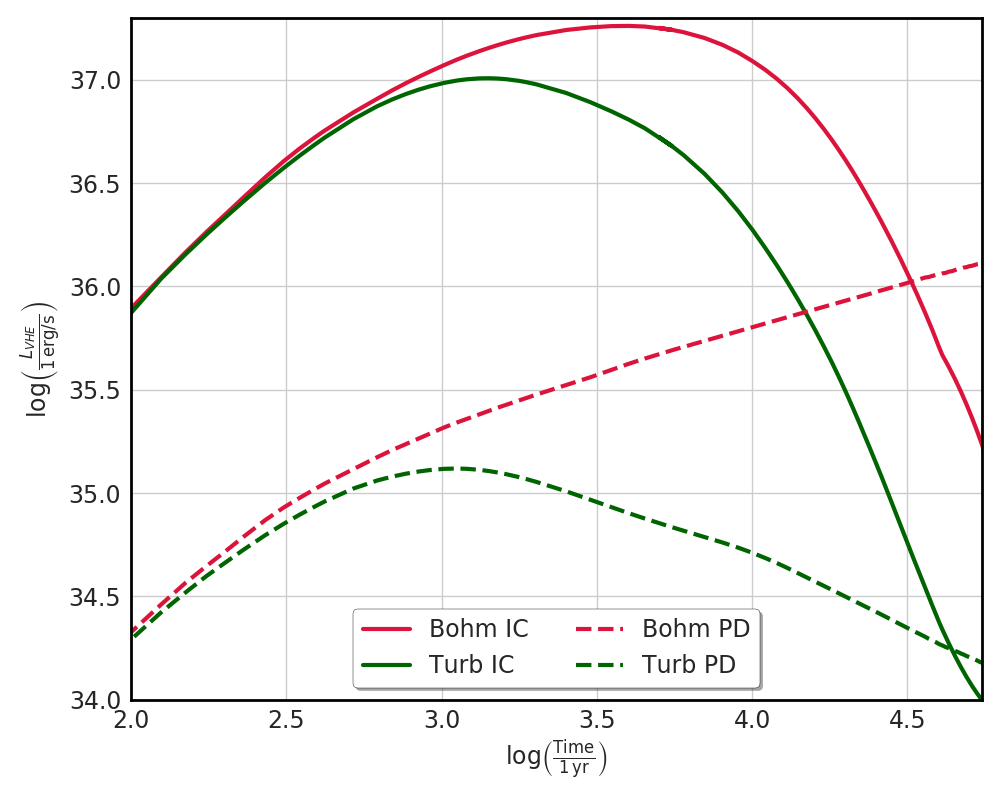}
\caption{Time evolution of the gamma-ray luminosity in the TeV band. The curves reflect models with Bohm-like (red) and Alfv\'enic (green) diffusion, and they indicate IC (solid lines) and PD (dashed lines) emission. The curves are calculated for a ambient gas density of $0.4\,\mathrm{cm}^{-3}$ assuming the same injection-efficiency for electrons and protons.}
\label{fig:GammaTheo}
\end{figure}

At least in the TeV band it is still unclear which emission process - inverse-Compton radiation or the decay of neutral pions - dominates the gamma-ray emission from SNRs. We evaluated the time-evolution of the VHE luminosity for both emission processes over the entire lifetime of an SNR and compared it to data of the H.E.S.S. galactic-SNR survey \citep{2018A&A...612A...3H}. Figure \ref{fig:GammaTheo} shows our prediction for the time evolution of the gamma-ray luminosity for inverse-Compton (IC) and Pion-decay (PD) emission in the $1-10\,$TeV band for Bohm-like and Alfv\'enic diffusion.    

Our models are calculated for a ambient density of $0.4\,\mathrm{cm}^ {-3}$ and assume the same injection efficiency for electrons and protons\footnote{The same injection efficiency will result in the same total number of electrons and protons in the simulation domain and an electron-to-proton ratio of $K_{ep}\simeq\sqrt{{m_e}/{m_p}}$ at relativistic energies \rbn{\citep{1993A&A...270...91P}}.}. \rbn{The contribution of background CRs is ignored in this section.}

There is a fundamental difference in the behaviour of IC and PD emission at later times that arises from the energy loss of electrons via synchrotron radiation, which suppresses the IC emission in the VHE-band after roughly $3000\,$yrs. The maximum energy of the electrons is approximately given by a balance between the energy loss rate and the acceleration rate,
\begin{equation}
    E_\mathrm{max} \approx 25\ \text{TeV}\, \left(\frac{v_\mathrm{sh}}{1000\,\text{km/s}}\right)\sqrt{\frac{5\,\mu\text{G}}{B_0}}\, ,\label{eq:Emax_e}
\end{equation}
where $m_e$, $q$, $v_sh$, and $B_0$ denote the electron mass, the elementary charge, the shock speed, and the upstream magnetic-field strength, respectively\footnote{The formula was derived assuming Bohm diffusion, a magnetic-field compression at the shock by $\sqrt{11}$ and only considering the upstream diffusion coefficient as important for the acceleration time.}. If the shock is slower than $2,000\,$km/s, the maximum electron energy falls below $50\,$TeV, and hence the IC-cutoff energy approaches $1\,$TeV. Our estimate suggests that the IC flux in the TeV band should start to decrease at an age of 3600, 1700 and 750 years for low ($n=0.04\,$cm$^{-3}$), medium ($n=0.4\,$cm$^{-3}$) and high ($n=4.0\,$cm$^{-3}$) density of the ambient medium, respectively. This roughly fits to our simulations in which the actual peak luminosity is reached later in all three cases as the expansion and thus the increase in the number of radiating particles initially compensates for the decreasing maximum energy.

Protons do not efficiently lose energy and can produce VHE gamma-rays throughout the lifetime of the SNR. Thus, the hadronic gamma-ray brightness keeps increasing with time for Bohm-like diffusion. In contrast, for Alfv\'enic diffusion the faster escape of CRs leads to a roll-off in the gamma-ray luminosity that is faster for leptonic emission and lower for hadronic channels but evident in both.  The luminosity peak is reached at earlier times than for Bohm-like diffusion, as any weaker driving of turbulence limits the acceleration efficiency.

\begin{figure*}[h]
\includegraphics[width=0.97\textwidth]{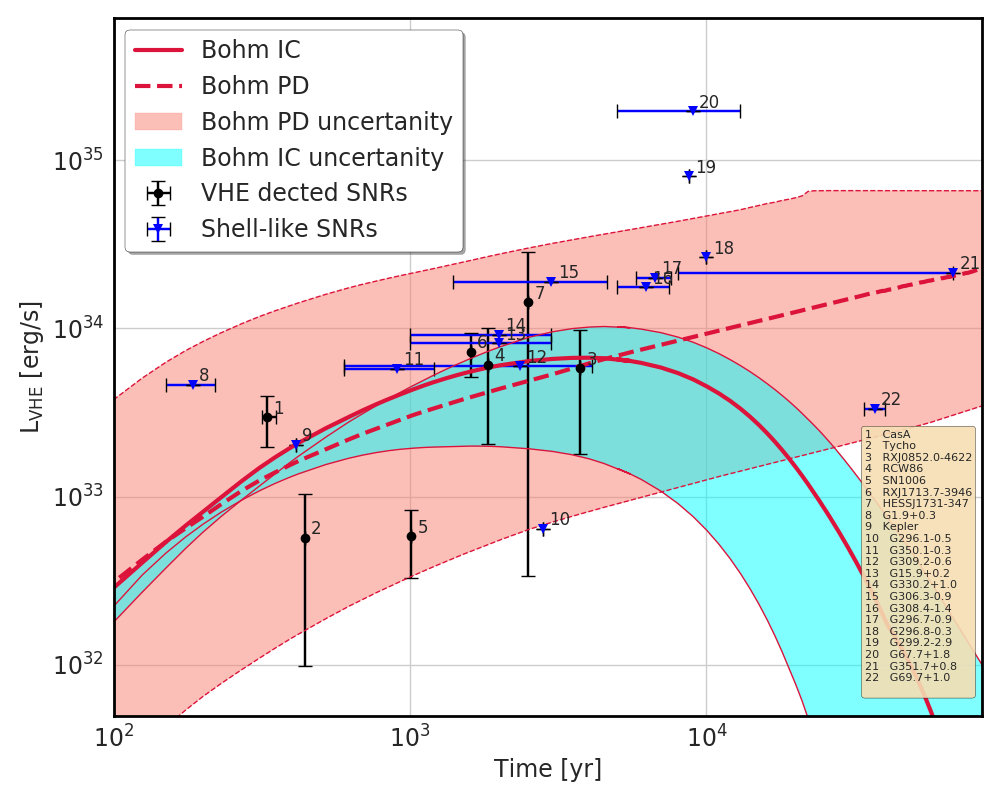}
\caption{The observed gamma-ray luminosity of galactic SNRs in the TeV band; blue markers indicate upper limits. The red curves reflect the models with Bohm-like diffusion and indicate inverse-Compton (solid lines) and pion-decay (dashed lines) emission. The thick curves are calculated for a ambient gas density of $0.4\,\mathrm{cm}^{-3}$. The shaded uncertainty bands represent the range of luminosity expected for a density in the range $0.04-4.0\,\mathrm{cm}^{-3}$. \rbn{The references for the data points are given in the text.}}
\label{fig:GammaAll}
\end{figure*}

The H.E.S.S. collaboration published a study examining the VHE luminosity of eight detected and several undetected SNRs with known distances, ages, and ambient densities \citep[and references therein]{2018A&A...612A...3H}\footnote{The flux for G69.7 is taken from \cite{2018ApJ...861..134A} and a conservative distance of $7.1\,$kpc was adopted \citep{2016ApJ...816....1K}.} In Figure \ref{fig:GammaAll} we compare our results for Bohm-like diffusions with these observations. To account for the different densities observed for these remnants, we ran two additional models with ambient densities of $0.04\,\mathrm{cm}^ {-3}$ and $4.0\,\mathrm{cm}^ {-3}$. All other parameters remained unchanged.

To reproduce the luminosity of the detected SNRs we had to lower the injection efficiency for electrons and protons, and we kept it the same  for all ambient densities. The electron to proton ratio is $0.025\sqrt{{m_e}/{m_p}}$ at high energies.

High ambient densities result - as expected - in a much brighter PD emission but suppress the IC emission at the same time on account of the rapid decline of the shock speed. Thus, the maximum electron energy starts to decrease already after one kyr as opposed to $2-4\,$kyrs in the medium and low-density cases. Low ambient densities thus allow for a higher IC peak luminosity as synchrotron cooling starts to decrease the maximum electron energy only after a larger total number of electrons got accelerated, compared to the medium and high-density cases.

Initially, the density-dependence of the IC emission is much weaker than that of the hadronic emission. The VHE luminosity is proportional to the number of CR particles, which for a fixed injection efficiency is only weakly dependent on the ambient density. In the adiabatic phase, one finds for the enclosed volume $V\propto \rho_0^{-3/5}$. The total number of particles injected into the acceleration process is thus $\propto \rho_0^{2/5}$. However, the PD-luminosity additionally depends on the ambient density, and the resulting scalings $L_{IC} \propto \rho_0^{2/5}$ and $L_{PD} \propto \rho_0^{7/5}$ are roughly observed in our simulations.

Our findings naturally explain why the brightest observed SNRs (Vela Jr, RXJ1713) expand in very low density environments and remnants with similar ages - like Puppis A, which expands in a much denser medium - have been missed by the IACTs \citep{2015A&A...575A..81H}.

We note that our predictions in Figure \ref{fig:GammaAll} are only valid up to SNR ages of roughly $4\,$kyrs as measurements indicate that the diffusion coefficients in the precursor of these remnants start to become higher than Bohm-like \citep{2018A&A...618A.155S}. In our Alfv\'enic scenario, we observe this transition already after about $2\,$kyrs. Afterwards, the escape of high-energy particles from the far-downstream region becomes important. As a consequence, the PD luminosity should start to fall as the particle spectra get softer at high energies (see Figure \ref{fig:GammaTheo}). We do observe a few bright SNR with ages around $10^4$~yrs. Their soft spectra are in line with the Alfv\'enic-turbulence model, but their fluxes fit better to the Bohm predictions. This indicates that interaction with dense clouds, additional turbulence amplification mechanisms, or other processes may play a significant role.


There are additional uncertainties like the filling factor or the type of supernova explosion that we have not taken into account in this simple model. For SN1006 it is for instance known that the non-thermal emission has a bipolar structure, indicating that just a fraction of the shock-surface is accelerating CRs. Thus, the brightness of a SNR can easily be a factor of a few below our model predictions. Further, most SNRs originate from core-collapse SNs and these remnants will, at least initially, expand in a density with a $r^{-2}$-dependence. Here, the total number of particles will be higher by about a factor three than predicted by measurements of the present-day post-shock density. For the density profiles
\begin{equation}
\rho(r) =\begin{cases}
\rho_u\qquad &\text{for Type 1a } \\
\rho_u\left({R_{sh}}/{r}\right)^2\qquad &\text{for CC}
\end{cases}
\end{equation}
the total number of particles passed through the shock is
\begin{align}
 N_\mathrm{tot} &= 4\pi\int \rho r^2 v_{sh} \D t\nonumber\\
 &= \begin{cases}
        \frac{4}{3}\pi R_{sh}^3 \rho_u \qquad &\text{for Type 1a } \\ 
        4\pi R_{sh}^3  \rho_u \qquad &\text{for CC } \ .
    \end{cases}\label{eq:Ntot}
\end{align}

We in detail discuss the differences between type-Ia and core-collapse SNR in the appendix \ref{sec:CoreCollapse}.
Similar variations arise when the shock recently encountered a density jump, like the edge of a wind-blown bubble. 
In that case the total number of particles will be lower than expected on the grounds of density measurements, and the remnant would consequently be dimmer. Kepler's SNR might be in this situation which indicates that dedicated modeling for each remnant is still needed.

\section{Conclusions}
We developed a model for the particle acceleration in SNRs up to a point where the forward shock of the remnant is radiative by solving the time-dependent transport equation for the CRs together with the gas-dynamical equations for the plasma flow. We investigated two diffusion regimes -- Bohm-like and Alfv\'enic diffusion. For the Alfv\'enic diffusion, we solved the time-dependent transport-equation for the evolution of the magnetic turbulence together with the gas-dynamical and the CR-equations. The ratio of CR to shock-ram pressure increased in all scenarios but remained well below 10\% at all times except for the very end of our Bohm-diffusion simulation. Hence our simulations were conducted in the test-particle limit. 

We showed that inefficient confinement in the Alfv\'enic diffusion regime of high-energy particles leads to a rapid reduction of the maximum energy reachable at the shock and to the eventual formation of a spectral break around $10-100\,$GeV. The spectra of CRs inside the SNR show no simple power-law structure above the break, but can be reasonably well approximated by a power law with an index of $s=2.7$. This spectral structure is similar to that observed from W44 and IC443.

The evaluation of the total production spectra, including CRs that reside outside of the SNR, showed also broken power-law spectra with a spectral index of $s\approx2.4$ at high energies. This is in rough agreement with the injection spectra required by galactic propagation models, namely around $s=2$ at low energies and a break to $s\simeq 2.4$ around $10$~GeV. The spectral index obtained for the release spectrum is thus harder than the spectrum of the CRs that still reside inside the SNR.

We investigated the reacceleration of galactic CRs at the shock. For efficient injection from the thermal pool, reaccelerated CRs would be too few to be noted, even for core-collapse supernovae. To render reaccelerated CRs visible, we artificially reduced the thermal injection of CRs at the shock. For an injection efficiency scaling as $\eta\propto t^{-1}$ galactic CRs became the dominant source for particles accelerated at the shock after $\approx2500\,$yrs. We showed that the resulting spectra above a few GeV are similar to those obtained by the acceleration of particles injected from the thermal pool. As a result, the emission signature at high energies is indistinguishable from that in the direct-acceleration scenario. However, the harder spectra of galactic electrons below a few GeV leave an imprint in the radio emission, as standard DSA can not soften spectra with an spectral index below $s=2$ but hardens spectra with a spectral index larger than $s=2$. Thus, the resulting radio spectra show a hardening below a few GHz - a feature that could be observable by radio and infrared telescopes.     
The emission resulting from CRs leaving the SNR and possibly illuminating a molecular cloud ahead of the shock shows narrow log-parabola spectra if the diffusion is Bohm-like but wide spectra if the diffusion is Alfv\'enic. Again, the difference between directly accelerated and reaccelerated CRs offers few, if any options to distinguish between both scenarios.

We examined the luminosity evolution of type-Ia SNRs assuming that the very-high energy gamma-rays originate either from pion decay or inverse-Compton radiation. For Bohm diffusion we found that the IC-luminosity is initially only weakly density dependent in contrast to the PD luminosity which scales roughly linearly with the ambient density. The total number of CRs accelerated in a SNR is proportional to the remnants size and the ambient density. However, the higher number of particles available for acceleration in high-density environments is largely compensated by the smaller size of the remnants compared to low-density environments. As a result, the density-dependence of the gamma-ray luminosity is roughly the same as the density dependence of the emission process itself.

In the case of Bohm-like diffusion, the remnant's IC luminosity peaks at an age around a few thousand years after which cooling suppresses the emission at high energies, while the PD luminosity keeps increasing with time. A study of different ambient densities revealed that SNRs expanding in low-density environments show a higher maximum IC-luminosity in accordance with observations - all the brightest VHE gamma-ray remnants RCW86, Vela Jr, HESSJ1721-347 and RXJ1713.7-3946 - expand in low-density environments. For a low ambient density the shock speed remains high for a long time and acceleration remains faster than synchrotron losses for electrons. 

If the diffusion coefficient is time-dependent as in the Alfv\'enic scenario, the PD-emission peaks as well. In this case both luminosities reach their maximum after roughly $1\,$kyr. Afterwards, cooling for electrons or the escape of high-energy protons reduce the VHE luminosity. This could explain why old SNRs are only detected in VHE gamma-ray when interacting with massive molecular clouds as the high target density for PD-emission significantly enhances their luminosity. However, the position of the peak indicates that either the interaction with the dense material began late in the evolution of the SNR or is otherwise restricted, or a more efficient amplification of turbulence needs to be considered.

This study and hence conclusions are based on the modeling of the type-Ia SNRs. However, we did model also core-collapse SNRs evolving in the wind-zone with the spatially-dependent magnetic field and found that results are comparable to type-Ia. The initially higher magnetic field might be important as it causes severe synchrotron losses but later on after around 500 years as the remnant expands into the shocked wind region the evolution of the core-collapse SNR basically mimics the evolution of the type-Ia SNR due to the roughly constant density and magnetic field.\\
\rb{\small \textit{Acknowledgements.} The authors gratefully thank the referee Mikhail Malkov for his constructive input.}

\bibliographystyle{aa}
\bibliography{References}

\begin{thebibliography}{68}
\expandafter\ifx\csname natexlab\endcsname\relax\def\natexlab#1{#1}\fi

\bibitem[{{Abdo} {et~al.}(2010){Abdo}, {Ackermann}, {Ajello}, {Allafort},
  {Baldini}, {Ballet}, {Barbiellini}, {Bastieri}, {Bechtol}, \&
  {Bellazzini}}]{2010ApJ...718..348A}
{Abdo}, A.~A., {Ackermann}, M., {Ajello}, M., {et~al.} 2010, \apj, 718, 348

\bibitem[{{Abeysekara} {et~al.}(2018){Abeysekara}, {Archer}, {Aune}, {Benbow},
  {Bird}, {Brose}, {Buchovecky}, {Bugaev}, {Cui}, {Daniel}, {Falcone}, {Feng},
  {Finley}, {Fleischhack}, {Flinders}, {Fortson}, {Furniss}, {Gotthelf},
  {Grube}, {Hanna}, {Hervet}, {Holder}, {Huang}, {Hughes}, {Humensky},
  {H{\"u}tten}, {Johnson}, {Kaaret}, {Kar}, {Kelley-Hoskins}, {Kertzman},
  {Kieda}, {Krause}, {Kumar}, {Lang}, {Lin}, {Maier}, {McArthur}, {Moriarty},
  {Mukherjee}, {O'Brien}, {Ong}, {Otte}, {Pandel}, {Park}, {Petrashyk}, {Pohl},
  {Popkow}, {Pueschel}, {Quinn}, {Ragan}, {Reynolds}, {Richards}, {Roache},
  {Rousselle}, {Rulten}, {Sadeh}, {Santander}, {Sembroski}, {Shahinyan},
  {Tyler}, {Vassiliev}, {Wakely}, {Ward}, {Weinstein}, {Wells}, {Wilcox},
  {Wilhelm}, {Williams}, \& {Zitzer}}]{2018ApJ...861..134A}
{Abeysekara}, A.~U., {Archer}, A., {Aune}, T., {et~al.} 2018, \apj, 861, 134

\bibitem[{{Acciari} {et~al.}(2009){Acciari}, {Aliu}, {Arlen}, {Aune},
  {Bautista}, {Beilicke}, {Benbow}, {Bradbury}, {Buckley}, {Bugaev}, {Butt},
  {Byrum}, {Cannon}, {Celik}, {Cesarini}, {Chow}, {Ciupik}, {Cogan}, {Colin},
  {Cui}, {Daniel}, {Dickherber}, {Duke}, {Dwarkadas}, {Ergin}, {Fegan},
  {Finley}, {Finnegan}, {Fortin}, {Fortson}, {Furniss}, {Gall}, {Gibbs},
  {Gillanders}, {Godambe}, {Grube}, {Guenette}, {Gyuk}, {Hanna}, {Hays},
  {Holder}, {Horan}, {Hui}, {Humensky}, {Imran}, {Kaaret}, {Karlsson},
  {Kertzman}, {Kieda}, {Kildea}, {Konopelko}, {Krawczynski}, {Krennrich},
  {Lang}, {LeBohec}, {Maier}, {McCann}, {McCutcheon}, {Millis}, {Moriarty},
  {Ong}, {Otte}, {Pandel}, {Perkins}, {Pohl}, {Quinn}, {Ragan}, {Reyes},
  {Reynolds}, {Roache}, {Rose}, {Schroedter}, {Sembroski}, {Smith}, {Steele},
  {Swordy}, {Theiling}, {Toner}, {Valcarcel}, {Varlotta}, {Vassiliev},
  {Vincent}, {Wagner}, {Wakely}, {Ward}, {Weekes}, {Weinstein}, {Weisgarber},
  {Williams}, {Wissel}, {Wood}, \& {Zitzer}}]{2009ApJ...698L.133A}
{Acciari}, V.~A., {Aliu}, E., {Arlen}, T., {et~al.} 2009, \apjl, 698, L133

\bibitem[{{Ackermann} {et~al.}(2013){Ackermann}, {Ajello}, {Allafort},
  {et~al.}}]{Fermi.2013a}
{Ackermann}, M., {Ajello}, M., {Allafort}, {et~al.} 2013, Science, 339, 807

\bibitem[{{Allen} {et~al.}(2008){Allen}, {Houck}, \&
  {Sturner}}]{2008ApJ...683..773A}
{Allen}, G.~E., {Houck}, J.~C., \& {Sturner}, S.~J. 2008, \apj, 683, 773

\bibitem[{{Bell} \& {Lucek}(2001)}]{2001MNRAS.321..433B}
{Bell}, A.~R. \& {Lucek}, S.~G. 2001, \mnras, 321, 433

\bibitem[{{Bell} {et~al.}(2019){Bell}, {Matthews}, \&
  {Blundell}}]{2019MNRAS.488.2466B}
{Bell}, A.~R., {Matthews}, J.~H., \& {Blundell}, K.~M. 2019, \mnras, 488, 2466

\bibitem[{{Bell} {et~al.}(2013){Bell}, {Schure}, {Reville}, \&
  {Giacinti}}]{2013MNRAS.431..415B}
{Bell}, A.~R., {Schure}, K.~M., {Reville}, B., \& {Giacinti}, G. 2013, \mnras,
  431, 415

\bibitem[{{Berezhko} {et~al.}(2003){Berezhko}, {Ksenofontov}, \&
  {V{\"o}lk}}]{2003A&A...412L..11B}
{Berezhko}, E.~G., {Ksenofontov}, L.~T., \& {V{\"o}lk}, H.~J. 2003, \aap, 412,
  L11

\bibitem[{{Blasi} {et~al.}(2005){Blasi}, {Gabici}, \& {Vannoni}}]{Blasi.2005a}
{Blasi}, P., {Gabici}, S., \& {Vannoni}, G. 2005, \mnras, 361, 907

\bibitem[{{Bohdan} {et~al.}(2019){Bohdan}, {Niemiec}, {Pohl}, {Matsumoto},
  {Amano}, \& {Hoshino}}]{2019ApJ...878....5B}
{Bohdan}, A., {Niemiec}, J., {Pohl}, M., {et~al.} 2019, \apj, 878, 5

\bibitem[{{Brose} {et~al.}(2016){Brose}, {Telezhinsky}, \&
  {Pohl}}]{2016A&A...593A..20B}
{Brose}, R., {Telezhinsky}, I., \& {Pohl}, M. 2016, \aap, 593, A20

\bibitem[{{Bykov} {et~al.}(2014){Bykov}, {Ellison}, {Osipov}, \&
  {Vladimirov}}]{2014ApJ...789..137B}
{Bykov}, A.~M., {Ellison}, D.~C., {Osipov}, S.~M., \& {Vladimirov}, A.~E. 2014,
  APJ, 789, 137

\bibitem[{{Caprioli} {et~al.}(2009){Caprioli}, {Blasi}, {Amato}, \&
  {Vietri}}]{Caprioli.2009}
{Caprioli}, D., {Blasi}, P., {Amato}, E., \& {Vietri}, M. 2009, MNRAS, 395, 895

\bibitem[{{Caprioli} {et~al.}(2018){Caprioli}, {Zhang}, \&
  {Spitkovsky}}]{2018JPlPh..84c7101C}
{Caprioli}, D., {Zhang}, H., \& {Spitkovsky}, A. 2018, Journal of Plasma
  Physics, 84, 715840301

\bibitem[{{Cardillo} {et~al.}(2016){Cardillo}, {Amato}, \&
  {Blasi}}]{2016A&A...595A..58C}
{Cardillo}, M., {Amato}, E., \& {Blasi}, P. 2016, \aap, 595, A58

\bibitem[{{Castelletti} {et~al.}(2007){Castelletti}, {Dubner}, {Brogan}, \&
  {Kassim}}]{2007A&A...471..537C}
{Castelletti}, G., {Dubner}, G., {Brogan}, C., \& {Kassim}, N.~E. 2007, \aap,
  471, 537

\bibitem[{{Celli} {et~al.}(2019){Celli}, {Morlino}, {Gabici}, \&
  {Aharonian}}]{2019arXiv190609454C}
{Celli}, S., {Morlino}, G., {Gabici}, S., \& {Aharonian}, F. 2019, arXiv
  e-prints, arXiv:1906.09454

\bibitem[{{Cox}(1972)}]{1972ApJ...178..159C}
{Cox}, D.~P. 1972, \apj, 178, 159

\bibitem[{{Cristofari} \& {Blasi}(2019)}]{2019MNRAS.489..108C}
{Cristofari}, P. \& {Blasi}, P. 2019, \mnras, 489, 108

\bibitem[{{Cui} {et~al.}(2018){Cui}, {Yeung}, {Tam}, \&
  {P{\"u}hlhofer}}]{2018ApJ...860...69C}
{Cui}, Y., {Yeung}, P. K.~H., {Tam}, P.~H.~T., \& {P{\"u}hlhofer}, G. 2018,
  \apj, 860, 69

\bibitem[{{Cummings} {et~al.}(2016){Cummings}, {Stone}, {Heikkila}, {Lal},
  {Webber}, {J{\'o}hannesson}, {Moskalenko}, {Orlando}, \&
  {Porter}}]{2016ApJ...831...18C}
{Cummings}, A.~C., {Stone}, E.~C., {Heikkila}, B.~C., {et~al.} 2016, \apj, 831,
  18

\bibitem[{{Diesing}(2019)}]{2019ICRC...36...59D}
{Diesing}, R. 2019, International Cosmic Ray Conference, 36, 59

\bibitem[{{Diesing} \& {Caprioli}(2019)}]{2019arXiv190507414D}
{Diesing}, R. \& {Caprioli}, D. 2019, arXiv e-prints, arXiv:1905.07414

\bibitem[{{Dwarkadas} \& {Chevalier}(1998)}]{1998ApJ...497..807D}
{Dwarkadas}, V.~V. \& {Chevalier}, R.~A. 1998, \apj, 497, 807

\bibitem[{{Gabici} \& {Aharonian}(2007)}]{2007ApJ...665L.131G}
{Gabici}, S. \& {Aharonian}, F.~A. 2007, \apjl, 665, L131

\bibitem[{{Giacalone} \& {Jokipii}(2007)}]{2007ApJ...663L..41G}
{Giacalone}, J. \& {Jokipii}, J.~R. 2007, \apjl, 663, L41

\bibitem[{{Giuliani} {et~al.}(2011){Giuliani}, {Cardillo}, {Tavani}, {Fukui},
  {Yoshiike}, {Torii}, {Dubner}, {Castelletti}, {Barbiellini}, \&
  {Bulgarelli}}]{2011ApJ...742L..30G}
{Giuliani}, A., {Cardillo}, M., {Tavani}, M., {et~al.} 2011, \apj, 742, L30

\bibitem[{{H.~E.~S.~S. Collaboration} {et~al.}(2018{\natexlab{a}}){H.~E.~S.~S.
  Collaboration}, {Abdalla}, {Abramowski}, {Aharonian}, {Ait Benkhali},
  {Akhperjanian}, {Andersson}, {Ang{\"u}ner}, {Arrieta}, {Aubert}, {Backes},
  {Balzer}, {Barnard}, {Becherini}, {Becker Tjus}, {Berge}, {Bernhard},
  {Bernl{\"o}hr}, {Blackwell}, {B{\"o}ttcher}, {Boisson}, {Bolmont}, {Bordas},
  {Bregeon}, {Brun}, {Brun}, {Bryan}, {Bulik}, {Capasso}, {Carr}, {Casanova},
  {Cerruti}, {Chakraborty}, {Chalme-Calvet}, {Chaves}, {Chen}, {Chevalier},
  {Chr{\'e}tien}, {Colafrancesco}, {Cologna}, {Condon}, {Conrad}, {Cui},
  {Davids}, {Decock}, {Degrange}, {Deil}, {Devin}, {deWilt}, {Dirson},
  {Djannati-Ata{\"\i}}, {Domainko}, {Donath}, {Drury}, {Dubus}, {Dutson},
  {Dyks}, {Edwards}, {Egberts}, {Eger}, {Ernenwein}, {Eschbach}, {Farnier},
  {Fegan}, {Fernand es}, {Fiasson}, {Fontaine}, {F{\"o}rster}, {Fukuyama},
  {Funk}, {F{\"u}{\ss}ling}, {Gabici}, {Gajdus}, {Gallant}, {Garrigoux},
  {Giavitto}, {Giebels}, {Glicenstein}, {Gottschall}, {Goyal}, {Grondin},
  {Hadasch}, {Hahn}, {Haupt}, {Hawkes}, {Heinzelmann}, {Henri}, {Hermann},
  {Hervet}, {Hinton}, {Hofmann}, {Hoischen}, {Holler}, {Horns}, {Ivascenko},
  {Jacholkowska}, {Jamrozy}, {Janiak}, {Jankowsky}, {Jankowsky}, {Jingo},
  {Jogler}, {Jouvin}, {Jung-Richardt}, {Kastendieck}, {Katarzy{\'n}ski},
  {Katz}, {Kerszberg}, {Kh{\'e}lifi}, {Kieffer}, {King}, {Klepser}, {Klochkov},
  {Klu{\'z}niak}, {Kolitzus}, {Komin}, {Kosack}, {Krakau}, {Kraus}, {Krayzel},
  {Kr{\"u}ger}, {Laffon}, {Lamanna}, {Lau}, {Lees}, {Lefaucheur}, {Lefranc},
  {Lemi{\`e}re}, {Lemoine-Goumard}, {Lenain}, {Leser}, {Lohse}, {Lorentz},
  {Liu}, {L{\'o}pez-Coto}, {Lypova}, {Marandon}, {Marcowith}, {Mariaud},
  {Marx}, {Maurin}, {Maxted}, {Mayer}, {Meintjes}, {Meyer}, {Mitchell},
  {Moderski}, {Mohamed}, {Mohrmann}, {Mor{\r{a}}}, {Moulin}, {Murach}, {de
  Naurois}, {Niederwanger}, {Niemiec}, {Oakes}, {O'Brien}, {Odaka}, {{\"O}ttl},
  {Ohm}, {Ostrowski}, {Oya}, {Padovani}, {Panter}, {Parsons}, {Pekeur},
  {Pelletier}, {Perennes}, {Petrucci}, {Peyaud}, {Piel}, {Pita}, {Poon},
  {Prokhorov}, {Prokoph}, {P{\"u}hlhofer}, {Punch}, {Quirrenbach}, {Raab},
  {Reimer}, {Reimer}, {Renaud}, {de los Reyes}, {Rieger}, {Romoli},
  {Rosier-Lees}, {Rowell}, {Rudak}, {Rulten}, {Sahakian}, {Salek}, {Sanchez},
  {Santangelo}, {Sasaki}, {Schlickeiser}, {Sch{\"u}ssler}, {Schulz},
  {Schwanke}, {Schwemmer}, {Settimo}, {Seyffert}, {Shafi}, {Shilon}, {Simoni},
  {Sol}, {Spanier}, {Spengler}, {Spies}, {Stawarz}, {Steenkamp}, {Stegmann},
  {Stinzing}, {Stycz}, {Sushch}, {Takahashi}, {Tavernet}, {Tavernier},
  {Taylor}, {Terrier}, {Tibaldo}, {Tiziani}, {Tluczykont}, {Trichard}, {Tuffs},
  {Uchiyama}, {van der Walt}, {van Eldik}, {van Rensburg}, {van Soelen},
  {Vasileiadis}, {Veh}, {Venter}, {Viana}, {Vincent}, {Vink}, {Voisin},
  {V{\"o}lk}, {Volpe}, {Vuillaume}, {Wadiasingh}, {Wagner}, {Wagner}, {Wagner},
  {White}, {Wierzcholska}, {Willmann}, {W{\"o}rnlein}, {Wouters}, {Yang},
  {Zabalza}, {Zaborov}, {Zacharias}, {Zdziarski}, {Zech}, {Zefi}, {Ziegler}, \&
  {{\.Z}ywucka}}]{2018A&A...612A...6H}
{H.~E.~S.~S. Collaboration}, {Abdalla}, H., {Abramowski}, A., {et~al.}
  2018{\natexlab{a}}, \aap, 612, A6

\bibitem[{{H.~E.~S.~S. Collaboration} {et~al.}(2018{\natexlab{b}}){H.~E.~S.~S.
  Collaboration}, {Abdalla}, {Abramowski}, {Aharonian}, {Ait Benkhali},
  {Ang{\"u}ner}, {Arakawa}, {Arrieta}, {Aubert}, \&
  {Backes}}]{2018A&A...612A...3H}
{H.~E.~S.~S. Collaboration}, {Abdalla}, H., {Abramowski}, A., {et~al.}
  2018{\natexlab{b}}, \aap, 612, A3

\bibitem[{{H.~E.~S.~S. Collaboration} {et~al.}(2015){H.~E.~S.~S.
  Collaboration}, {Abramowski}, {Aharonian}, {Ait Benkhali}, {Akhperjanian},
  {Ang{\"u}ner}, {Backes}, {Balenderan}, {Balzer}, \&
  {Barnacka}}]{2015A&A...575A..81H}
{H.~E.~S.~S. Collaboration}, {Abramowski}, A., {Aharonian}, F., {et~al.} 2015,
  \aap, 575, A81

\bibitem[{{Hanusch} {et~al.}(2019){Hanusch}, {Liseykina}, \&
  {Malkov}}]{2019ApJ...872..108H}
{Hanusch}, A., {Liseykina}, T.~V., \& {Malkov}, M. 2019, \apj, 872, 108

\bibitem[{{Jaffe} {et~al.}(2011){Jaffe}, {Banday}, {Leahy}, {Leach}, \&
  {Strong}}]{2011MNRAS.416.1152J}
{Jaffe}, T.~R., {Banday}, A.~J., {Leahy}, J.~P., {Leach}, S., \& {Strong},
  A.~W. 2011, \mnras, 416, 1152

\bibitem[{{Kilpatrick} {et~al.}(2016){Kilpatrick}, {Bieging}, \&
  {Rieke}}]{2016ApJ...816....1K}
{Kilpatrick}, C.~D., {Bieging}, J.~H., \& {Rieke}, G.~H. 2016, \apj, 816, 1

\bibitem[{{Li} {et~al.}(2018){Li}, {Zhou}, {Huang}, {Zhang}, {Qiao}, {Yu},
  {Ruan}, \& {He}}]{2018PhPl...25h2103L}
{Li}, R., {Zhou}, C.~T., {Huang}, T.~W., {et~al.} 2018, Physics of Plasmas, 25,
  082103

\bibitem[{{L{\'o}pez-Coto} \& {Giacinti}(2018)}]{2018MNRAS.479.4526L}
{L{\'o}pez-Coto}, R. \& {Giacinti}, G. 2018, \mnras, 479, 4526

\bibitem[{{Lucek} \& {Bell}(2000)}]{2000MNRAS.314...65L}
{Lucek}, S.~G. \& {Bell}, A.~R. 2000, \mnras, 314, 65

\bibitem[{{Malkov}(1998)}]{1998PhRvE..58.4911M}
{Malkov}, M.~A. 1998, \pre, 58, 4911

\bibitem[{{Malkov} \& {Aharonian}(2019)}]{2019ApJ...881....2M}
{Malkov}, M.~A. \& {Aharonian}, F.~A. 2019, \apj, 881, 2

\bibitem[{{Malkov} {et~al.}(2011){Malkov}, {Diamond}, \&
  {Sagdeev}}]{Malkov.2011a}
{Malkov}, M.~A., {Diamond}, P.~H., \& {Sagdeev}, R.~Z. 2011, Nature
  Communications, 2, 194

\bibitem[{{Malkov} {et~al.}(2012){Malkov}, {Diamond}, \&
  {Sagdeev}}]{2012PhPl...19h2901M}
{Malkov}, M.~A., {Diamond}, P.~H., \& {Sagdeev}, R.~Z. 2012, Physics of
  Plasmas, 19, 082901

\bibitem[{{Matsumoto} {et~al.}(2017){Matsumoto}, {Amano}, {Kato}, \&
  {Hoshino}}]{Matsumoto2017}
{Matsumoto}, Y., {Amano}, T., {Kato}, T.~N., \& {Hoshino}, M. 2017, Phys. Rev.
  Lett.

\bibitem[{{Mignone} {et~al.}(2007){Mignone}, {Bodo}, {Massaglia}, {Matsakos},
  {Tesileanu}, {Zanni}, \& {Ferrari}}]{2007ApJS..170..228M}
{Mignone}, A., {Bodo}, G., {Massaglia}, S., {et~al.} 2007, \apjs, 170, 228

\bibitem[{{Moskalenko} \& {Strong}(1998)}]{1998ApJ...493..694M}
{Moskalenko}, I.~V. \& {Strong}, A.~W. 1998, \apj, 493, 694

\bibitem[{{Moskalenko} {et~al.}(2002){Moskalenko}, {Strong}, {Ormes}, \&
  {Potgieter}}]{2002ApJ...565..280M}
{Moskalenko}, I.~V., {Strong}, A.~W., {Ormes}, J.~F., \& {Potgieter}, M.~S.
  2002, \apj, 565, 280

\bibitem[{{Ohira} {et~al.}(2010){Ohira}, {Murase}, \&
  {Yamazaki}}]{2010A&A...513A..17O}
{Ohira}, Y., {Murase}, K., \& {Yamazaki}, R. 2010, \aap, 513, A17

\bibitem[{{Orlando}(2018)}]{2018MNRAS.475.2724O}
{Orlando}, E. 2018, \mnras, 475, 2724

\bibitem[{{Petruk} \& {Kopytko}(2016)}]{2016MNRAS.462.3104P}
{Petruk}, O. \& {Kopytko}, B. 2016, \mnras, 462, 3104

\bibitem[{{Petruk} {et~al.}(2016){Petruk}, {Kuzyo}, \&
  {Beshley}}]{2016MNRAS.456.2343P}
{Petruk}, O., {Kuzyo}, T., \& {Beshley}, V. 2016, \mnras, 456, 2343

\bibitem[{{Pohl}(1993)}]{1993A&A...270...91P}
{Pohl}, M. 1993, \aap, 270, 91

\bibitem[{{Ptuskin} \& {Zirakashvili}(2003)}]{2003A&A...403....1P}
{Ptuskin}, V.~S. \& {Zirakashvili}, V.~N. 2003, \aap, 403, 1

\bibitem[{{Ptuskin} \& {Zirakashvili}(2005)}]{2005A&A...429..755P}
{Ptuskin}, V.~S. \& {Zirakashvili}, V.~N. 2005, \aap, 429, 755

\bibitem[{{Putze} {et~al.}(2009){Putze}, {Derome}, {Maurin}, {Perotto}, \&
  {Taillet}}]{2009A&A...497..991P}
{Putze}, A., {Derome}, L., {Maurin}, D., {Perotto}, L., \& {Taillet}, R. 2009,
  \aap, 497, 991

\bibitem[{{Reville} \& {Bell}(2013)}]{2013MNRAS.430.2873R}
{Reville}, B. \& {Bell}, A.~R. 2013, \mnras, 430, 2873

\bibitem[{{Simpson} {et~al.}(2016){Simpson}, {Pakmor}, {Marinacci}, {Pfrommer},
  {Springel}, {Glover}, {Clark}, \& {Smith}}]{2016ApJ...827L..29S}
{Simpson}, C.~M., {Pakmor}, R., {Marinacci}, F., {et~al.} 2016, \apj, 827, L29

\bibitem[{{Skilling}(1975)}]{Skilling.1975a}
{Skilling}, J. 1975, MNRAS, 172, 557

\bibitem[{{Sushch} {et~al.}(2018){Sushch}, {Brose}, \&
  {Pohl}}]{2018A&A...618A.155S}
{Sushch}, I., {Brose}, R., \& {Pohl}, M. 2018, \aap, 618, A155

\bibitem[{{Sutherland} \& {Dopita}(1993)}]{1993ApJS...88..253S}
{Sutherland}, R.~S. \& {Dopita}, M.~A. 1993, \apjs, 88, 253

\bibitem[{{Tang} \& {Chevalier}(2015)}]{2015ApJ...800..103T}
{Tang}, X. \& {Chevalier}, R.~A. 2015, \apj, 800, 103

\bibitem[{{Telezhinsky} {et~al.}(2012{\natexlab{a}}){Telezhinsky}, {Dwarkadas},
  \& {Pohl}}]{Telezhinsky.2012a}
{Telezhinsky}, I., {Dwarkadas}, V.~V., \& {Pohl}, M. 2012{\natexlab{a}},
  Astroparticle Physics, 35, 300

\bibitem[{{Telezhinsky} {et~al.}(2012{\natexlab{b}}){Telezhinsky}, {Dwarkadas},
  \& {Pohl}}]{Telezhinsky.2012b}
{Telezhinsky}, I., {Dwarkadas}, V.~V., \& {Pohl}, M. 2012{\natexlab{b}}, A\&A,
  541, A153

\bibitem[{{Telezhinsky} {et~al.}(2013){Telezhinsky}, {Dwarkadas}, \&
  {Pohl}}]{Telezhinsky.2013}
{Telezhinsky}, I., {Dwarkadas}, V.~V., \& {Pohl}, M. 2013, \aap, 552, A102

\bibitem[{{Trotta} {et~al.}(2011){Trotta}, {J{\'o}hannesson}, {Moskalenko},
  {Porter}, {Ruiz de Austri}, \& {Strong}}]{2011ApJ...729..106T}
{Trotta}, R., {J{\'o}hannesson}, G., {Moskalenko}, I.~V., {et~al.} 2011, \apj,
  729, 106

\bibitem[{{van Marle} {et~al.}(2018){van Marle}, {Casse}, \&
  {Marcowith}}]{2018MNRAS.473.3394V}
{van Marle}, A.~J., {Casse}, F., \& {Marcowith}, A. 2018, \mnras, 473, 3394

\bibitem[{{V{\"o}lk} {et~al.}(2003){V{\"o}lk}, {Berezhko}, \&
  {Ksenofontov}}]{2003A&A...409..563V}
{V{\"o}lk}, H.~J., {Berezhko}, E.~G., \& {Ksenofontov}, L.~T. 2003, A\&A, 409,
  563

\bibitem[{{Yang} {et~al.}(2014){Yang}, {de O{\~n}a Wilhelmi}, \&
  {Aharonian}}]{2014A&A...566A.142Y}
{Yang}, R.-z., {de O{\~n}a Wilhelmi}, E., \& {Aharonian}, F. 2014, \aap, 566,
  A142

\bibitem[{{Zeng} {et~al.}(2019){Zeng}, {Xin}, \& {Liu}}]{2019ApJ...874...50Z}
{Zeng}, H., {Xin}, Y., \& {Liu}, S. 2019, \apj, 874, 50

\bibitem[{{Zirakashvili} \& {Ptuskin}(2012)}]{2012APh....39...12Z}
{Zirakashvili}, V.~N. \& {Ptuskin}, V.~S. 2012, Astroparticle Physics, 39, 12

\end{thebibliography}

\appendix

\section{Core-collapse SNRs}\label{sec:CoreCollapse}
We also modeled the acceleration of CRs at a core-collapse SN expanding in a wind zone, using the initial conditions described in \cite{2018A&A...618A.155S} and a magnetic field that is $\propto 1/r$.  The parameters are given in Table \ref{table:RSGpars}. They are roughly consistent with those expected for the environment of red-supergiants (RSG). After $\approx 1800\,$yrs the density and magnetic-field strength are comparable to those used for the medium-density type-Ia simulations.

\begin{table}[h]
	\centering
	\caption{}
	\label{table:RSGpars}
	\begin{tabular}{@{}lcc@{}}
		\hline
Parameter &   RSG & Type 1a\\
		\hline
       $E_\mathrm{ej}$ [erg] & $10^{51}$ & $10^{51}$ \\
       $M_\mathrm{ej}$ [$\text{M}_\odot$] & 3 & 1.4 \\
       $v_\mathrm{wind}$ [km/s] & 20 & - \\
       $\dot{M_\star}$ [$\text{M}_\odot$/yr] & $10^{-4}$ & - \\
       t [yr] & 1800 & 1800 \\
       $R_\mathrm{sh}(t)$ [pc] & 5.5 & 7.9 \\
       $v_\mathrm{sh}(t)$ [km/s] & 2150 & 1900 \\
       $n_u(t)$ [cm$^{-3}$] & 0.4 & 0.4 \\
       $B_u(t)$ [$\mu$G] & 5 & 5 \\
		\hline
	\end{tabular}
\end{table}

\begin{figure}[h]
\includegraphics[width=0.5\textwidth]{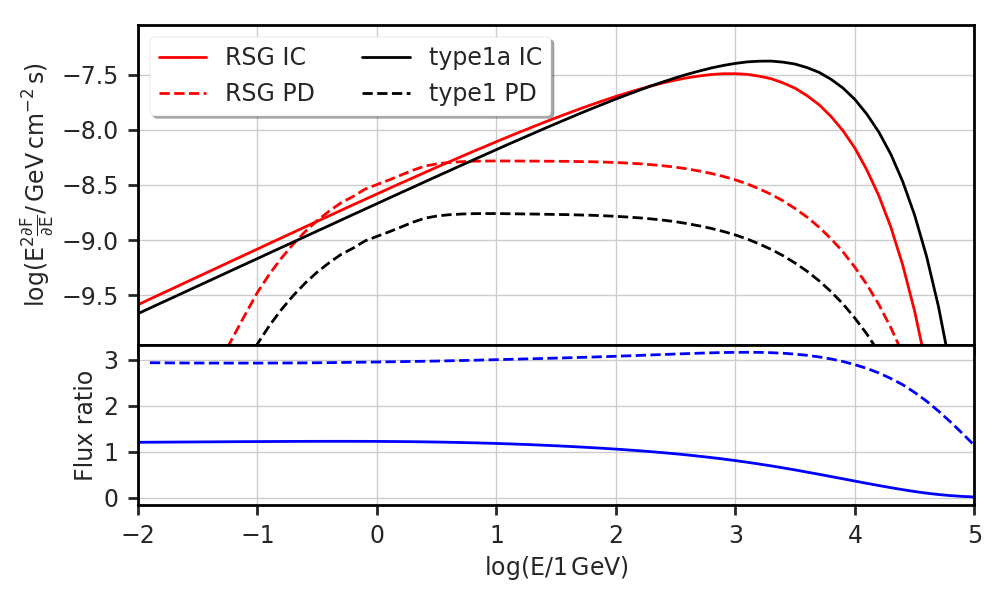}
\caption{Top panel: Gamma-ray spectra of IC (solid) and PD (dashed) emission for a RSG (red) and a type-Ia (black) SNR at 1800 years.
Bottom panel: Gamma-ray flux ratio of an RSG vs. a type-Ia environment for PD (dashed) and IC (solid) radiation.}
\label{fig:GammaCC}
\end{figure}

The remnant size for a RSG is typically smaller than that for a type-Ia explosion, as the material around the star is initially much denser. However, the shock speed is comparable after 1800~years, and we calculated the gamma-ray emission for that time. The spectra are presented in Figure \ref{fig:GammaCC}.

It can be seen, that the PD gamma-ray flux is about three times higher in the CC-case compared to the type-Ia scenario. Although the volume of a type-Ia SNR is roughly three times larger than that of its core-collapse sibling, equation \ref{eq:Ntot} indicates that both the number of cosmic rays and the total amount of target material are increased by a factor three, hence the overall gain by a factor three. 
For similar reasons the IC gamma-ray intensities are comparable between CC and type-Ia SNRs as the differences caused by the size and the number of cosmic rays cancel out. The initially higher magnetic field in the RSG-scenario lowers the maximum energy of electrons because synchrotron cooling is more efficient (see also Equation \ref{eq:Emax_e}). As a result, the leptonic gamma-ray luminosity in the $1-10\,$TeV band is about six times lower in the RSG case than that in the type-Ia scenario. This indicates that the spatial structure around the SNR is one of the factors that determine the IC gamma-ray luminosity. All the above applies only as long as the remnant expands in the unshocked wind. As soon as the shock enters the shocked wind, it is again propagating in an area of approximately constant density. This transition usually happens earlier -- after around $500\,$ years for a RSG -- than the age of $1800\,$ years we are considering here. After entering the shocked wind, the impact of the initial wind-zone will fade and our solutions for type-Ia will become valid again, because core-collapse and type-Ia explosion have a comparable explosion energy, which is -- besides the ambient density -- the crucial parameter that defines the evolution during the Sedov stage. A detailed modeling of SNRs expanding in the various environments around massive stars is important to understand the gamma-ray emission of early SNRs but that is clearly beyond the scope of this paper.

The role of background CRs is still negligible in the RSG scenario, even after the remnant entered the low-density environment of the shocked wind. In this case, the contribution of background CRs can be of the order of $1\,$\% when effects of stellar modulation are neglected. In reality, low-energy CRs will have been pushed out of the circumstellar medium by the wind of the progenitor star, and the background level of CRs will be even lower.

\section{Alternative CR background spectrum}\label{sec:AlternateBackground}

\begin{figure}[h]
\includegraphics[width=0.5\textwidth]{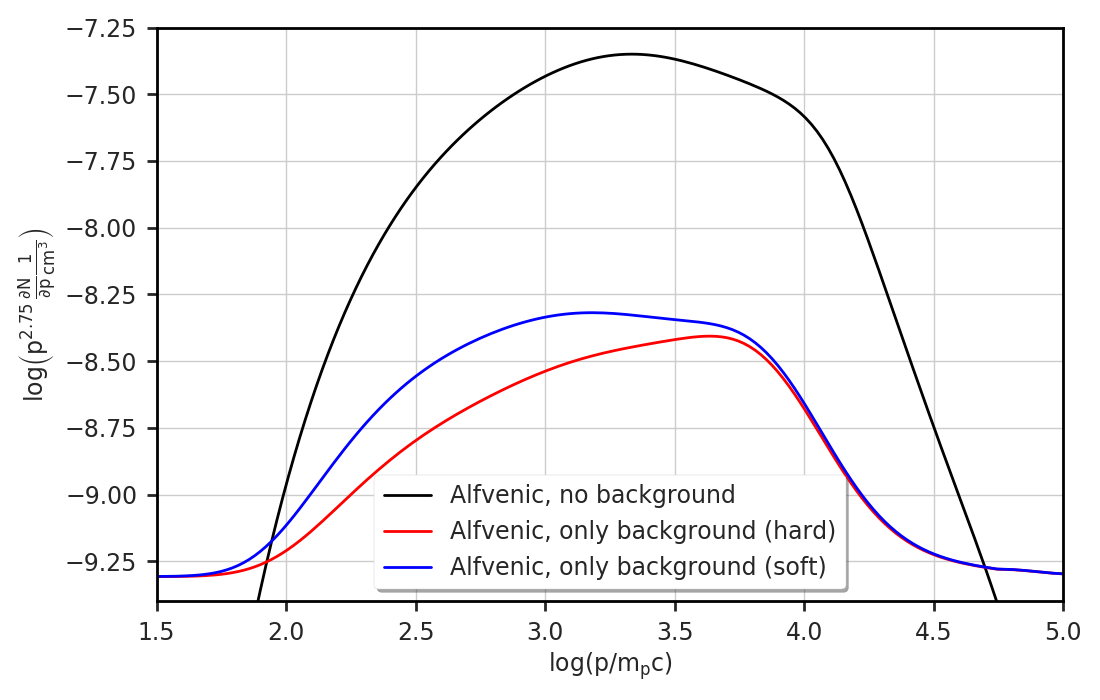}
\caption{CR  spectra  about  $7.5\,$pc  ahead  of  the  forward
shock at an age of 37 kyrs.  The black line refers to the acceleration of particles from the thermal pool, where the red (blue) line refer the reacceleration of the hard (soft) background spectrum below $1\,$GeV. Note that the y-axis is scaled with $p^{2.75}$.}
\label{fig:AltBack}
\end{figure}

The background CR spectrum given in equation (\ref{Eq:BackgroundPRs}) is rather hard below $1$~GeV, and it has been argued that the true spectrum might be softer \citep{2018MNRAS.475.2724O}. There has to be some low-energy cutoff, otherwise the energy deposited in low-energetic cosmic rays would not be finite, but the position of this cutoff is unknown. We investigated to what extend our spectra are affected by our choice of the background CR spectrum. We repeated our reacceleration runs with the alternative background spectrum 
\begin{align}
    N_\mathrm{PR}(E) &= 
        \begin{cases}
            N_0 E^{-2.75}\,\, \text{ for } E\geq\text{1GeV} \\
            N_0 E^{-1.5}\,\,\, \text{ for } E<\text{1GeV} 
        \end{cases} .\label{Eq:BackgroundPRs_2}
\end{align}
As a result, more low-energetic CRs are present to drive the amplification of magnetic turbulence and the confinement of CRs around the SNR. The difference in the spectra is seen most significantly in the escape spectra. Figure \ref{fig:AltBack} compares the volume-averaged spectra in a shell $7.5\,$pc ahead of the forward shock after $37\,$kyrs.

The higher number of low-energy particles enables a stronger driving of turbulence and a better confinement of these particles around the SNR. As a result, the escape spectrum is softer for the softer CR-background spectrum and approaches the spectral index of the CRs accelerated from the thermal pool of particles. In this sense, our original background spectrum represents a lower limit to number of low-energy particles that are available to drive the magnetic turbulence and confine the CRs in and around the SNR. The softer the CR spectrum below $\approx1\,$GeV, the closer the resemblance of the reacceleration spectra to those arising from acceleration from the thermal pool.  

\end{document}